\documentclass[prx,twocolumn,superscriptaddress,citeautoscript,showpacs,amsart]{revtex4}

\usepackage{graphicx}
\usepackage{multirow}
\usepackage{color}
\usepackage{bm}
\usepackage{times}
\usepackage{amsmath,bm,amsfonts}
\usepackage{dcolumn}
\usepackage{graphicx}
\usepackage{latexsym}

\usepackage{ulem} 
\usepackage{braket}

\def\ket#1{|#1\rangle }

\def\bb{\mathbb}
\def\Pf{{\rm Pf}}

\begin{document}

\title{Unconventional topological phase transition in two-dimensional systems
with space-time inversion symmetry}

\author{Junyeong \surname{Ahn}}
\affiliation{Department of Physics and Astronomy, Seoul National University, Seoul 08826, Korea}

\affiliation{Center for Correlated Electron Systems, Institute for Basic Science (IBS), Seoul 08826, Korea}

\affiliation{Center for Theoretical Physics (CTP), Seoul National University, Seoul 08826, Korea}

\author{Bohm-Jung \surname{Yang}}
\email{bjyang@snu.ac.kr}
\affiliation{Department of Physics and Astronomy, Seoul National University, Seoul 08826, Korea}

\affiliation{Center for Correlated Electron Systems, Institute for Basic Science (IBS), Seoul 08826, Korea}

\affiliation{Center for Theoretical Physics (CTP), Seoul National University, Seoul 08826, Korea}

\date{\today}

\begin{abstract}
We study a topological phase transition between a normal insulator and a quantum spin Hall insulator in two-dimensional (2D) systems with time-reversal and two-fold rotation symmetries. 
Contrary to the case of ordinary time-reversal invariant systems where a direct transition between two insulators is generally predicted, we find that the topological phase transition in systems with an additional two-fold rotation symmetry is mediated by an emergent stable two-dimensional Weyl semimetal phase between two insulators. 
Here the central role is played by the so-called space-time inversion symmetry, the combination of time-reversal and two-fold rotation symmetries, which guarantees the quantization of the Berry phase around a 2D Weyl point even in the presence of strong spin-orbit coupling.
Pair-creation/pair-annihilation of Weyl points accompanying partner exchange between different pairs induces a jump of a 2D $Z_{2}$ topological invariant leading to a topological phase transition. 
According to our theory, the topological phase transition in HgTe/CdTe quantum well structure is mediated by a stable 2D Weyl semimetal phase since the quantum well, lacking inversion symmetry intrinsically, has two-fold rotation about the growth direction. Namely, the HgTe/CdTe quantum well can show 2D Weyl semimetallic behavior within a small but finite interval in the thickness of HgTe layers between a normal insulator and a quantum spin Hall insulator. We also propose that few-layer black phosphorus under perpendicular electric field is another candidate system to observe the unconventional topological phase transition mechanism accompanied by emerging 2D Weyl semimetal phase protected by space-time inversion symmetry.
\end{abstract}


\maketitle

{\it Introduction.$-$}
Symmetry protected topological phases have become a quintessential notion
in condensed matter physics, after the discovery of time-reversal invariant topological insulators~\cite{Hasan-Kane,Qi-Zhang,Ando-Fu} and topological crystalline insulators~\cite{TCI}. 
Although the importance of symmetry to protect bulk topological properties
is widely recognized, relatively little attention has been paid to understand the role of symmetry for the description of topological phase transition (TPT).
Early studies on this issue have focused on time-reversal $T$ and inversion $P$, and have shown that the nature of TPT in $T$-invariant three-dimensional (3D) systems changes dramatically depending on the presence or absence of $P$ symmetry~\cite{Murakami}. 
Namely, when the system has both $T$ and $P$ symmetries, a direct transition between a normal insulator (NI) and a $Z_{2}$ topological insulator is possible when a band inversion happens between two bands with opposite parities.
Whereas in noncentrosymmetric systems lacking $P$, the transition between a NI and a topological insulator is generally mediated by a 3D Weyl semimetal (WSM) phase in between. 
The intermediate stable semimetal phase can appear when the following two conditions are satisfied. Firstly, the codimension analysis for accidental band crossing at a generic momentum should predict a group of gapless solutions. Secondly, a gapless point in the semimetal phase should carry a quantized topological invariant guaranteeing its stability.

\begin{figure}[b]
\centering
\includegraphics[width=8.5 cm]{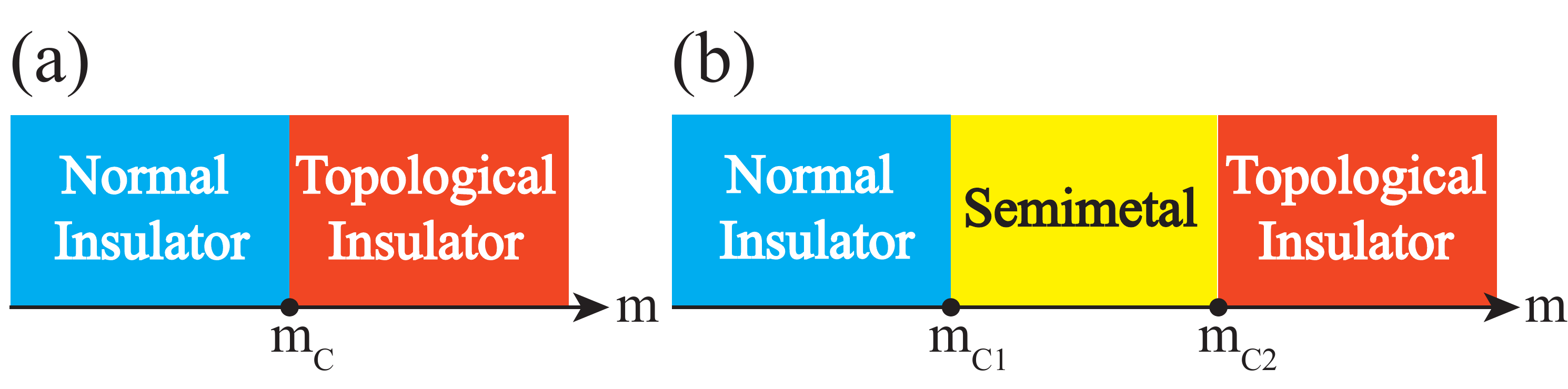}
\caption{
Schematic phase diagram for a topological phase transition in 
a time-reversal invariant 2D noncentrosymmetric system.
(a) For systems only with time-reversal symmetry.
(b) For systems with an additional two-fold rotation symmetry about an axis
perpendicular to the 2D plane.
}
\label{phasediagram}
\end{figure}

Contrary to 3D, in two-dimensions (2D), the codimension analysis~\cite{Murakami_2D} predicts that there always is a direct transition between a NI and a quantum spin Hall insulator (QSHI) even in noncentrosymmetric systems (See Fig.~\ref{phasediagram}(a)). The reason is that, in a generic 2D system with a single tuning parameter $m$ (representing pressure, doping, etc.), an effective $2\times 2$ Hamiltonian describing band crossing depends on three independent variables $(k_{x},k_{y},m)$ including two momenta $k_{x}$ and $k_{y}$. To achieve a band crossing, however, since the coefficients of three Pauli matrices associated with the effective $2\times 2$ Hamiltonian should vanish by adjusting three variables, only a single gap-closing solution can be found. This unique gap-closing solution describes the critical point for a direct transition between two gapped insulators.  
The absence of a stable semimetal phase mediating the transition between two insulators
is consistent with the fact that a gap-closing point does not carry a topological invariant in a generic $T$-invariant 2D system in the presence of spin-orbit coupling.

In this Letter, we show that the TPT
between a NI and a QSHI is always mediated by an emerging 2D Weyl~\cite{notation} semimetal (See Fig.~\ref{phasediagram}(b)), when a $T$-invariant noncentrosymmetric 2D system is invariant under two-fold rotation $C_{2z}$ about an axis perpendicular to the 2D plane. The intermediate 2D WSM is stable due to the $\pi$ Berry phase around a Weyl point (WP), which is quantized even in the presence of
spin-orbit coupling.
Here the central role is played by the so-called space-time inversion $I_{ST}$ which is nothing but the combination of $T$ and $C_{2z}$, i.e., $I_{ST}=C_{2z}T$.
Since $I_{ST}$ ensures the quantization of the Berry phase around a 2D WP, the transition between an insulator and a WSM is accompanied by pair-creation and pair-annihilation of 2D WPs.
Moreover, partner exchange between pairs of WPs can induce the change
of the $Z_{2}$ topological invariant, thus the 2D WSM can mediate a TPT. 
We propose two candidate materials where the unconventional TPT mediated by a 2D WSM can be realized. One is the HgTe/CdTe quantum well where inversion is absent intrinsically~\cite{BIA_Dai,BIA_Konig,BIA_Winkler,BIA_Zunger} and the TPT can be controlled
by changing the thickness of the HgTe layer. We expect that there can be a finite thickness window where a stable 2D WSM appears between a NI and a QSHI.
Also we propose that the unconventional TPT can be observed in
few-layer black phosphorus under vertical electric field.

{\it Band crossing in systems with $I_{ST}$.$-$} 
$C_{2z}$ transforms a spatial coordinate $(x,y,z)$ to $(-x,-y,z)$. Since the $z$-coordinate is invariant under $C_{2z}$, for a layered 2D system with a fixed $z$, $C_{2z}$ can be considered as
an effective inversion symmetry mapping $(x,y)$ to $(-x,-y)$.
Thus, in 2D systems,
$I_{ST}=C_{2z}T$ transforms a space-time coordinate $(x,y,t)$ to $(-x,-y,-t)$. In momentum space, on the other hand, it is a local symmetry since the momentum $\bm{k}=(k_{x},k_{y})$ remains invariant under $I_{ST}$.
As discussed in Ref.~\onlinecite{spacetime}, $I_{ST}$
has various intriguing properties.
For instance, since Berry curvature $F_{xy}(\bm{k})$ transforms to $-F_{xy}(\bm{k})$ under $I_{ST}$, $F_{xy}(\bm{k})$ vanishes locally unless there is a singular gapless point, which guarantees the quantization of $\pi$ Berry phase around a 2D WP.
Moreover, since $I_{ST}^{2}=+1$ irrespective of the presence/absence of spin-orbit coupling, it does not require Kramers degeneracy at each $\bm{k}$.
This can be contrasted to the case of $PT$, satisfying $(PT)^{2}=-1~(+1)$
in the presence (absence) of spin-orbit coupling. Especially, when $(PT)^{2}=-1$, Kramers theorem requires double degeneracy at each $\bm{k}$. Since Berry phase is not quantized in this case, a Dirac point is unstable~\cite{Kane-Mele}.

Here we show that $I_{ST}$ also modifies the gap-closing condition in an essential way, leading to an unconventional TPT.
Since each band is nondegenerate at a generic momentum $\bm{k}$ in noncentrosymmetric systems, an accidental band crossing can be described by a $2\times 2$ matrix Hamiltonian
$H(\bm{k})=f_{0}(\bm{k},m)+\sum_{i=x,y,z}f_{i}(\bm{k},m)\sigma_{i}$ where $\sigma_{x,y,z}$ indicates the two bands touching near the Fermi level and $m$ describes
a tuning parameter such as electric field, pressure, etc.
Since $I_{ST}$ is antiunitary, it can generally be represented by $I_{ST}=UK$ where $U$ is a unitary matrix and $K$ denotes complex conjugation.
By choosing a suitable basis, one can obtain $I_{ST}=K$ as shown in Supplemental Materials. Then the $I_{ST}$ symmetry requires $H^{*}(\bm{k})=H(\bm{k})$, which leads to
\begin{align}
H(\bm{k})=f_{0}(\bm{k},m)+f_{x}(\bm{k},m)\sigma_{x}+f_{z}(\bm{k},m)\sigma_{z}.
\end{align}
Accidental gap-closing can happen if and only if $f_{x}=f_{z}=0$. Since there are three
independent variables $(k_{x},k_{y},m)$ whereas there are only two equations $f_{x}=f_{z}=0$ to be satisfied, one can expect a line of gapless solutions in $(k_{x},k_{y},m)$ space, which predicts an emerging 2D stable semimetal (See Fig.~\ref{phasediagram}(b)). Near the critical point $m=m_{c1}$ where accidental band crossing happens, the Hamiltonian can generally be written as
\begin{align}\label{eqn:criticalHamiltonian}
H(\bm{q})=(Aq_{x}^{2}+m_{c1}-m)\sigma_{x}+vq_{y}\sigma_{z},
\end{align}
which describes a gapped insulator (a 2D WSM) when $m<m_{c1}$ ($m>m_{c1}$) assuming $A>0$. 
Due to $T$ symmetry, accidental band crossing happens at two momenta $\pm\bm{k}$. Since two WPs are created at each band crossing point, the WSM has four WPs in total. 
Moreover, when $m$ becomes larger than $m_{c1}$, four WPs migrate in momentum space, and eventually, they are annihilated pairwise at $m=m_{c2}$.  Interestingly, when pair-annihilation/pair-annihilation is accompanied by partner-switching between WP pairs, the two gapped phases mediated by the WSM should have
distinct topological property as shown below.

\begin{figure}[h!]
\centering
\includegraphics[width=8.5 cm]{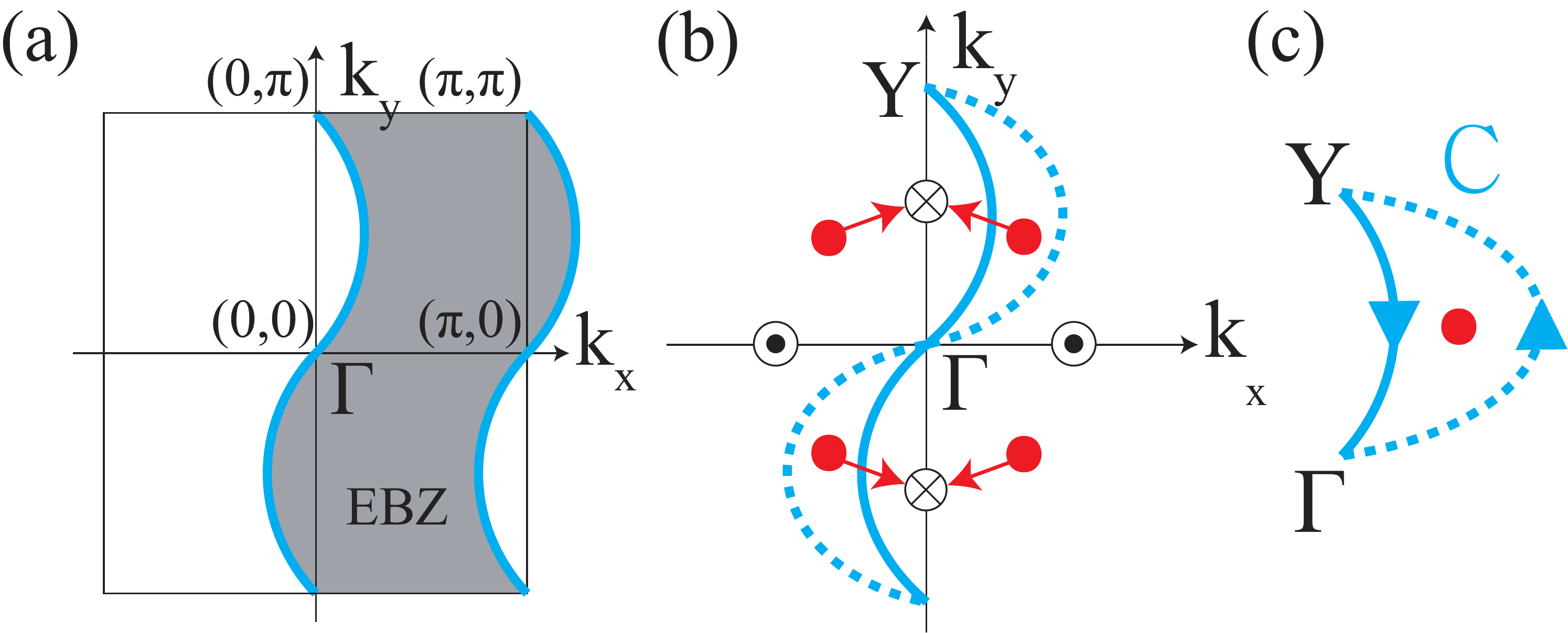}
\caption{(a) Two blue lines denote 1D $T$-invariant subsystems 
passing two time-reversal invariant
momenta with $k_{x}=0$ or $k_{x}=\pi$, respectively.
EBZ indicates the half Brillouin zone bounded by these two $T$-invariant 1D subsystems.
(b) A schematic figure describing the motion of Weyl points (WPs)
and the associated change in the topological invariant of a $T$-invariant 1D subsystem.
Red dots indicate 2D WPs whose trajectories are described by
red arrows. The solid (dotted) line indicates the 1D subsystem before (after) a gap-closing due to the relevant WPs.   
$\odot$ and $\otimes$ are locations where pair-creation and pair-annihilation of WPs happen.  
(c) A closed loop $C$ encircling a 2D WP.
}
\label{deformed_1d}
\end{figure}

{\it Change of $Z_{2}$ invariant via pair-creation/pair-annihilation of 2D WPs.$-$} 
The $Z_{2}$ invariant $\Delta$ of a $T$-invariant 2D system can be written as~\cite{Fu-Kane}
\begin{align}\label{eqn:Z2definition}
\Delta=P_{T}(\pi)-P_{T}(0)\quad \text{mod}~~2
\end{align}
where $P_{T}(k_{x})$ is the time-reversal polarization
of a $T$-invariant one-dimensional (1D) subsystem connecting two time-reversal invariant
momenta (TRIM) with given $k_{x}=0,\pi$. For instance, Fig.~\ref{deformed_1d}(a) shows $T$-invariant 1D subsystems passing two TRIMs with $k_{x}=0$ or $k_{x}=\pi$, respectively. For such a 1D subsystem, $P_{T}$ is defined as
\begin{align}
P_{T}=P^{I}-P^{II}=2P^{I}-P_{\rho},
\end{align}
where $P_{\rho}=P^{I}+P^{II}$ is the charge polarization, and
$P^{I}$ and $P^{II}$ are the partial polarization associated with 
the wave function $u^{I}_{n}(k)$ and its Kramers partner $u^{II}_{n}(-k)\propto Tu^{I}_{n}(k)$ where $n$ labels occupied bands. 
Namely, $P^{j=I,II}=\oint\frac{dk}{2\pi} A^{j}(k)$ with $A^{j}(k)=i\sum_{n}\langle u^{j}_{n}(k)|\nabla_{k}|u^{j}_{n}(k)\rangle$. 
$P^{j=I,II}$ can also be written as a summation of Wannier function centers such as $P^{j=I,II}=\sum_{n}~^{j}\langle R,n|r|R,n\rangle^{j}$ using the Wannier function $|R,n\rangle^{j}=\int\frac{dk}{2\pi} e^{-ik(R-r)}|u^{j}_{n}(k)\rangle$~\cite{Fu-Kane, Wannier1, Wannier2}.
In general, $P_{\rho}$ can take any real value, modulo an integer, whereas $P_{T}$ is an integral quantity whose magnitude is gauge dependent.
Thus in a generic 1D $T$-invariant system, among $P_{T}$, $P_{\rho}$, $2P^{I}$, none of them can serve as a topological invariant.

However, in the presence of additional $C_{2z}$ symmetry, 
$2P^{I}=P_{T}+P_{\rho}$ becomes a $Z_{2}$ topological invariant~\cite{supp}.
Namely, $2P^{I}$ becomes quantized and gauge-invariant modulo 2 when $I_{ST}$ exists. Since $C_{2z}$ makes $P^{I}$ to take a quantized value, either $0$
or $\frac{1}{2}$ modulo an integer, $2P^{I}$ naturally becomes a $Z_{2}$ quantity.
$2P^{I}$ is also proposed as a $Z_{2}$ topological invariant in a $T$-invariant 1D system
with mirror symmetry in Ref.~\onlinecite{mirror_insulator}.

Now let us explain how the pair-creation/pair-annihilation of 2D WPs
can change the $Z_{2}$ invariant $\Delta$ given by
\begin{align}
\Delta=2P^{I}(\pi)-2P^{I}(0)-\left(P_{\rho}(\pi)-P_{\rho}(0)\right).
\end{align}
In an insulating phase, since the Chern number of the whole system is zero,
one can choose a continuous gauge in which
\begin{align}
&P_{\rho}(\pi)-P_{\rho}(0)=\int_{\text{EBZ}}d^{2}k F(\bm{k})=0,
\end{align}
where EBZ indicates the effective half-Brillouin zone bounded by
two $T$-invariant 1D systems defined above, and we have used that Berry curvature
$F(\bm{k})=0$ due to $I_{ST}$.
Thus the change of $\Delta$ is simply given by $\delta\Delta=\delta\left(2P^{I}(\pi)\right)-\delta\left(2P^{I}(0)\right)$.
Since $2P^{I}$ is a topological invariant, it can be changed only
if an accidental gap-closing happens in the relevant 1D $T$-invariant subsystem. 
In the process shown in Fig.~\ref{deformed_1d}(b) where 2D WPs pass through
the 1D subsystem with $k_{x}=0$, we have
\begin{align}
\delta\Delta=-\delta\left(2P^{I}(0)\right)
=-\frac{1}{\pi}\oint_{C}d\bm{k}\cdot\bm{A}(\bm{k})
\end{align}
where $C$ is a closed loop encircling one WP shown in Fig.~\ref{deformed_1d}(c).
Since a 2D WP has $\pi$-Berry phase, $\delta\Delta=1$ ($\delta\Delta=0$)
if $C$ encloses an odd (even) number of WPs. Hence the
$Z_{2}$ invariant $\Delta$ can be changed by 1 via partner exchange between two
pairs of 2D WPs.

\begin{figure}[t]
\centering
\includegraphics[width=8.5 cm]{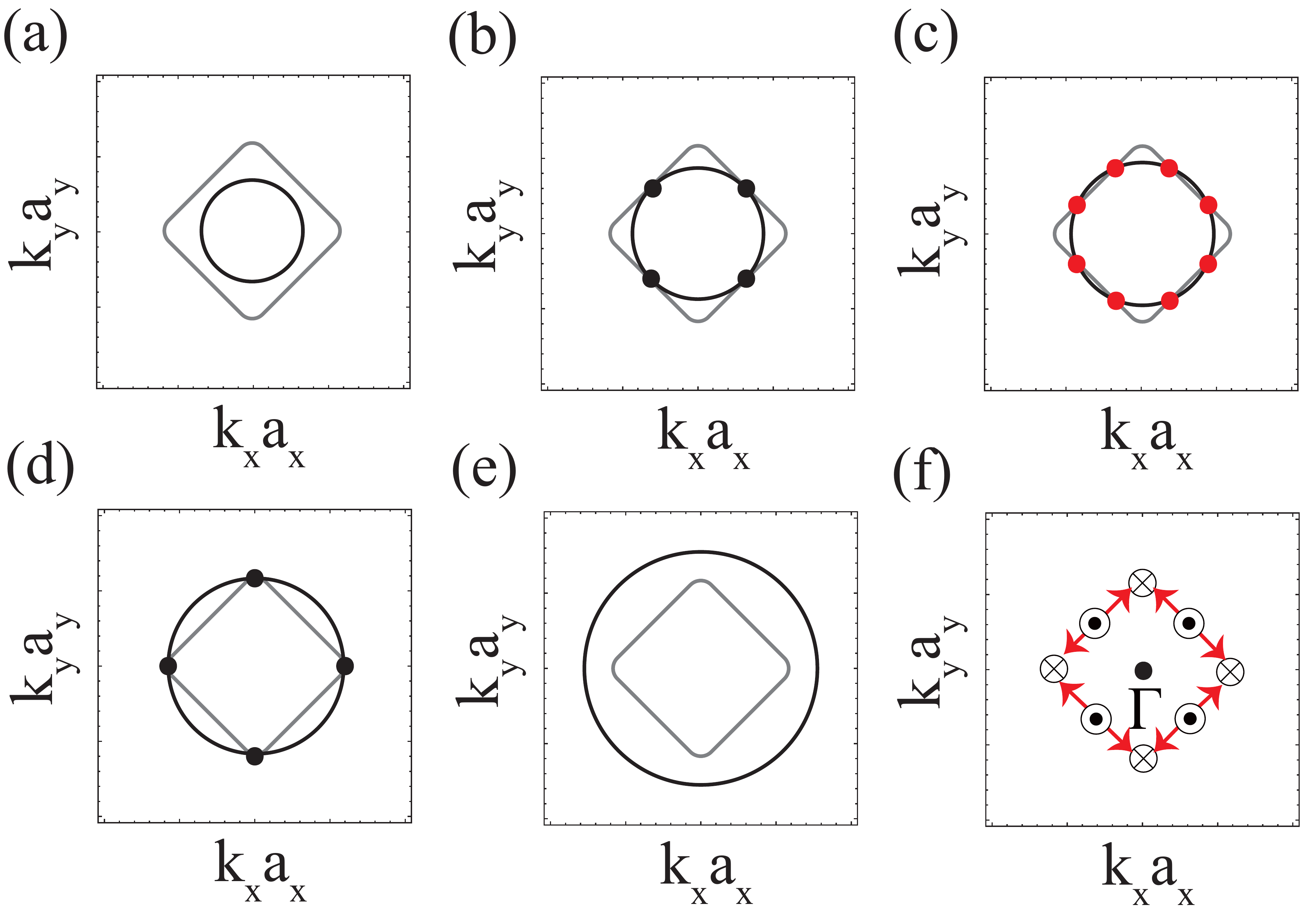}
\caption{(Color online)
Schematic figures describing the TPT in the BHZ model including an inversion breaking term.
Two loops in each panel indicate
the momenta where two gap-closing conditions,
$d_{1}^{2}+d_{2}^{2}=\lambda^{2}$ (grey line) and $d_{3}=0$ (black line) are satisfied, respectively.
(a) When $\frac{M}{B}<\left(\frac{M}{B}\right)_{c1}$ relevant to a NI.
(b) When $\frac{M}{B}=\left(\frac{M}{B}\right)_{c1}$.
(c) When $\left(\frac{M}{B}\right)_{c1}<\frac{M}{B}<\left(\frac{M}{B}\right)_{c2}$
relevant to a WSM. Here each red dot indicates a Weyl point (WP).
(d) When $\frac{M}{B}=\left(\frac{M}{B}\right)_{c2}$.
(e) When $\frac{M}{B}>\left(\frac{M}{B}\right)_{c2}$ relevant to a QSHI.
(f) The momentum space trajectory of WPs as $\frac{M}{B}$ increases
from $\left(\frac{M}{B}\right)_{c1}$ to $\left(\frac{M}{B}\right)_{c2}$.
Here $\odot$ and $\otimes$ are the locations where pair-creation and pair-annihilation of WPs happen,
respectively.
}
\label{fig:tight-binding}
\end{figure}

{\it Model Hamiltonian.$-$}
To demonstrate the unconventional TPT, we construct
a simple model Hamiltonian, a variant of Bernevig-Hughes-Zhang (BHZ) model,
which is originally proposed to describe QSH effect in a HgTe/CdTe quantum well~\cite{BHZ,BHZexp}.
The BHZ model is defined on a square lattice
in which each site has two s-orbitals $|s,\uparrow\rangle$, $|s,\downarrow\rangle$
and two spin-orbit coupled p-orbitals $|p_{x}+i p_{y},\uparrow\rangle$,
$|p_{x}-i p_{y},\downarrow\rangle$. 
Nearest-neighbor hopping between these four orbitals gives a tight-binding Hamiltonian
given by
\begin{align}
H_{\text{BHZ}}(\bm{k})=\varepsilon(\bm{k})+d_{1}(\bm{k})\sigma_{x}s_{z}
+d_{2}(\bm{k})\sigma_{y}+d_{3}(\bm{k})\sigma_{z},
\end{align}
where $d_{1}(\bm{k})+id_{2}(\bm{k})=A[\sin k_{x}+i\sin k_{y}]$,
$d_{3}(\bm{k})=-2B[2-\frac{M}{2B}-\cos k_{x}-\cos k_{y}]$,
$\varepsilon(\bm{k})=C-2D[2-\cos k_{x}-\cos k_{y}]$, and the Pauli matrices $\sigma_{x,y,z}$ ($s_{x,y,z}$) denote the orbital (spin) degrees of freedom.
$H_{\text{BHZ}}$ describes a QSHI (a NI)
when $0<\frac{M}{2B}<2$ ($\frac{M}{2B}<0$). 
The TPT can also be understood from
the energy eigenvalues of $H_{\text{BHZ}}$, $E(\bm{k})=\varepsilon(\bm{k})\pm\sqrt{d_{1}^{2}(\bm{k})+d_{2}^{2}(\bm{k})+d_{3}^{2}(\bm{k})}$.
The energy gap closes when three equations $d_{1}=d_{2}=d_{3}=0$
are simultaneously satisfied, which uniquely determines the three variable $(k_{x},k_{y},\frac{M}{B})=(0,0,0)$
when $\frac{M}{B}<4$.

$H_{\text{BHZ}}$ is invariant under inversion $P=\sigma_{z}$, four-fold rotation
about $z$ axis $C_{4z}=(\sigma_{z}+is_{z})/\sqrt{2}$, two-fold rotation 
about $x$, $y$ axis $C_{2x}=i\sigma_{z}s_{x}$, $C_{2y}=is_{y}$, and time-reversal $T=is_{y}K$.
However, the real HgTe system lacks $P$ and $C_{4z}$ while retaining their product $PC_{4z}^{3}\equiv S_{4}$ symmetry as well as $T$, $C_{2x}$, $C_{2y}$, and $C_{2z}=C_{4z}^{2}$. In fact, once $P$ is absent, a constant term $\lambda\sigma_{y}s_{y}$ is allowed. Then the resulting Hamiltonian
$H_{\text{HgTe}}(\bm{k})=H_{\text{BHZ}}(\bm{k})+\lambda\sigma_{y}s_{y}$
is invariant under $T$, $C_{2x}$, $C_{2y}$, $C_{2z}$, and $S_{4}$.
The energy eigenvalues of $H_{\text{HgTe}}(\bm{k})$ are
$E(\bm{k})=\varepsilon(\bm{k})\pm\sqrt{(\sqrt{d_{1}^{2}+d_{2}^{2}}\pm|\lambda|)^{2}+d_{3}^{2}}$.
Since gap-closing requires only two conditions, $d_{1}^{2}+d_{2}^{2}=\lambda^{2}$ and $d_{3}=0$ to be satisfied, one can expect a line of gapless solutions in $(k_{x},k_{y},\frac{M}{B})$ space describing a WSM. 
Generally, the solution of each gap-closing condition forms a closed loop in momentum space. 
In Fig.~\ref{fig:tight-binding}, we plot the evolution in the shape of two loops which describe the momenta
satisfying $d_{1}^{2}+d_{2}^{2}=\lambda^{2}$ and $d_{3}=0$, respectively.
Only when these two loops overlap, the band gap closes at the momentum where two loops touch, thus the system becomes a WSM. 
We find that,
when $\left(\frac{M}{B}\right)_{c1}<\frac{M}{B}<\left(\frac{M}{B}\right)_{c2}$,
two loops overlap at eight points indicating an emergent WSM having 8 WP.
Here $\left(\frac{M}{B}\right)_{c1}\equiv4-\sqrt{16-2\left(\frac{2\lambda}{A}\right)^{2}}$
and $\left(\frac{M}{B}\right)_{c2}\equiv2-\sqrt{4-\left(\frac{2\lambda}{A}\right)^{2}}$.
On the other hand, when $\frac{M}{B}<\left(\frac{M}{B}\right)_{c1}$ or
$\frac{M}{B}>\left(\frac{M}{B}\right)_{c2}$, two loops do not overlap, thus
the sytem is a gapped insulator.
At the critical point with $\frac{M}{B}=\left(\frac{M}{B}\right)_{c1,c2}$, the band gap closes
at four points related by $S_{4}$, and each splits into two WPs in the WSM phase. This result demonstrates that {\it the TPT in a HgTe/CdTe quantum well can be mediated by an intermediate 2D WSM.}
The occurrence of a semimetal phase in the HgTe/CdTe quantum well is also proposed in Ref.~\onlinecite{BIA_Winkler} based on symmetry analyses.
Let us note that the number of gap-closing points at the critical point depends
on the symmetry of the system. For instance, once $S_{4}$ symmetry
is broken by applying uniaxial strain, one can observe two gap-closing points
at the critical point and the intermediate WSM has four WPs. (See Supplemental Materials.)
However, irrespective of the number of gap-closing points, the WSM
can mediate a TPT as long as the trajectory of WPs forms a single closed loop.

To prove that the 2D WSM mediates a TPT,
we should compare the $Z_{2}$ invariant $\Delta$ of two insulating phases
existing when $\frac{M}{B}<\left(\frac{M}{B}\right)_{c1}$
and $\frac{M}{B}>\left(\frac{M}{B}\right)_{c2}$, respectively.
For this purpose, we first compute the energy spectrum of a strip structure
having a finite size along
one direction.
As shown in Fig.~\ref{fig:phase_transition} (b), when $\frac{M}{B}>\left(\frac{M}{B}\right)_{c2}$, one can clearly observe helical edge states
localized on the sample boundary, which is absent when $\frac{M}{B}<\left(\frac{M}{B}\right)_{c1}$.
Thus the system is a QSHI (a NI) when $\frac{M}{B}>\left(\frac{M}{B}\right)_{c2}$ ($\frac{M}{B}<\left(\frac{M}{B}\right)_{c1}$).
For further confirmation, we directly compute $\Delta$ numerically.
Since inversion symmetry is broken, one cannot use parity eigenvalues to evaluate $\Delta$~\cite{Fu-Kane_inversion}. Instead we determine $\Delta$ by computing the change of the time-reversal polarization $P_{T}$ between $k_{x}=0$ and $k_{x}=\pi$ by using Eq.~(\ref{eqn:Z2definition}). Since $P_{T}$ is given by the difference in the Wannier function centers of Kramers pairs, one can determine $\Delta$ by examining how Wannier function centers of Kramers pairs evolve between $k_{x}=0$ and $k_{x}=\pi$~\cite{Fu-Kane,wilson_loop}. As shown in Fig.~\ref{fig:phase_transition} (c),
when $\frac{M}{B}>\left(\frac{M}{B}\right)_{c2}$ ($\frac{M}{B}<\left(\frac{M}{B}\right)_{c1}$), one can see partner-switching (no partner-switching) of Wannier functions when $k_{x}$ changes from 0 to $\pi$.
Since the partner-switching (no partner-switching) between Wannier states indicates the change of
$P_{T}$ by 1 (0), one obtains
$\Delta=1$ ($\Delta=0$) when $\frac{M}{B}>\left(\frac{M}{B}\right)_{c2}$ ($\frac{M}{B}<\left(\frac{M}{B}\right)_{c1}$).


\begin{figure}[t]
\centering
\includegraphics[width=8.5cm]{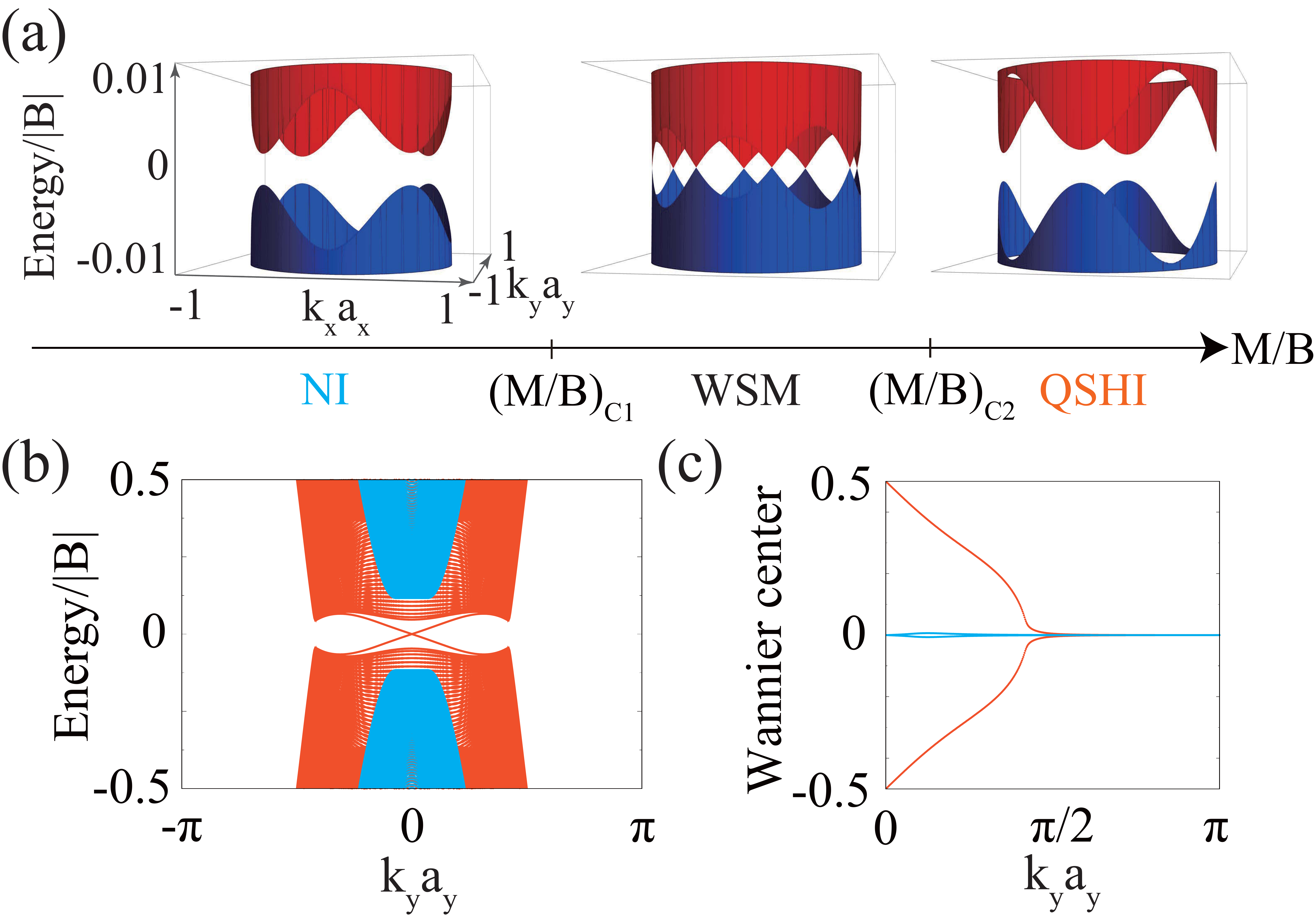}
\caption{(Color online)
(a) Evolution of the band structure across the TPT obtained from the BHZ model
including the inversion breaking term with $A/B=0.2$, $\lambda/B=0.15$.
To clarify the band structure near the gap-closing point, we set $\varepsilon(\bm{k})=0$.
The representative band structures are calculated at $M/B=0.595$ (NI), $0.645$ (WSM) and $0.695$ (QSHI).
(b) Energy spectrum of a finite-size strip structure.
(c) The evolution of the Wannier function centers.
In (b,c) blue and orange lines are relevant to when $M/B=0$ (NI) and $M/B=1.5$ (QSHI), respectively. 
}
\label{fig:phase_transition}
\end{figure}

{\it Discussion.$-$}
The unconventional TPT mediated by a WSM can generally occur
in any 2D noncentrosymmetric system with $I_{ST}$ symmetry.
Similar to the BHZ model including an inversion breaking term,
we have found that the Kane-Mele model on the honeycomb lattice including Rashba coupling also undergoes a TPT mediated by a 2D WSM when uniaxial strain is applied (See Supplemental Materials).
Among real materials, we propose few-layer black phosphorus
as an another candidate system
since its band gap can be controlled by electric field breaking inversion~\cite{phosphorene}.
Although there are several theoretical proposal for possible TPT
in this system~\cite{Zunger,Zhang,Sisakht},
the unconventional mechanism we propose has never been discussed.
In fact, a recent experiment~\cite{KSKim} has shown that the band gap of this system can be controlled by doping potassium on the surface, thus the insulator-semimetal transition
can be realized. Also the presence of 2D WPs in the semimetal phase
is observed in a first-principles calculation~\cite{Choi}. We expect, when stronger electric field is applied to the WSM phase, even a QSHI can be obtained.
For confirmation, we have studied a tight-binding model describing black phosphorus under vertical electric field, and have shown that the unconventional TPT can occur. (See Supplemental Materials for details.)
Since the presence of $C_{2z}$ with broken inversion is the only requirement to realize the novel TPT, the same phenomenon may occur in various 2D materials with $C_{2z}$ such as the puckered honeycomb structure of arsenene, antimonene, bismuthene~\cite{2Dmaterial},
the dumbbell structure of germanium-tin, stanene~\cite{dumbbell}, and the bismuth monobromide~\cite{monobromide}, etc.

We conclude with the discussion about electron correlation and disorder effects
on the TPT.
Although 2D WSM is perturbatively stable against weak interaction, sufficiently strong interaction can induce nontrivial physical consequences. For instance, a phase
breaking $T$ symmetry is proposed to appear between a NI and a QSHI due to interaction~\cite{interaction}.
Especially, at the critical point between the WSM and an insulator,
since the energy dispersion becomes anisotropic, i.e., linear in one direction and
quadratic in the other direction (See Eq.~(\ref{eqn:criticalHamiltonian})), the density of states shows $D(E)\propto\sqrt{E}$ with the energy $E$, contrary to $D(E)\propto E$ in the WSM with linear dispersion in two directions. 
Such an enhancement of $D(E)$ makes the electron
correlation and disorder to cause nontrivial physical consequences.
For instance, a recent renormalization group study~\cite{RG-largeN}
has shown that quantum fluctuation of anisotropic Weyl fermions makes
the screened Coulomb interaction to have spatial anisotropy, 
which eventually leads to marginal Fermi liquid behavior of 
low energy quasi-particles.
Also, in the presence of disorder, a disorder-induced
new semimetal phase can appear between the insulator and WSM~\cite{disorder}.
Understanding the interplay of electron correlation and disorder is an important problem, which we leave for future study.

{\it Acknowledgement.$-$} 
J. Ahn was supported by IBS-R009-D1.
B.-J. Y was supported by IBS-R009-D1, Research Resettlement Fund for the new faculty of Seoul National University, and Basic Science Research Program through the National Research Foundation of Korea (NRF) funded by the Ministry of Education (Grant No. 0426-20150011). We thank Keunsu Kim, Akira Furusaki, Takahiro Morimoto for useful discussion.



\renewcommand{\thefigure}{S\arabic{figure}}
\renewcommand{\theequation}{S\arabic{equation}}
\renewcommand{\thesection}{SI \arabic{section}}
\renewcommand{\thesubsection}{\Alph{subsection}}

\setcounter{figure}{0} 
\setcounter{equation}{0} 
\setcounter{section}{0} 

\section{Derivation of $I_{ST}=K$}

In this section, we show that $I_{ST}$ can be expressed as the complex conjugation $K$ via a unitary transformation of the basis .
Since space-time inversion is antiunitary, in general, it can be written as
\begin{align}
I_{ST}=U_{ST}K
\end{align}
where $U_{ST}$ is a unitary operator.
From the property $I_{ST}^2=1$ we see that $U_{ST}$ is symmetric.
Then $U_{ST}$ can be written in the form of
\begin{align}
U_{ST}=e^{iM},
\end{align}
where $M$ is real and symmetric.
The unitary part of space-time inversion transform as 
\begin{align}U_{ST}\rightarrow U^{\dagger}U_{ST}U^*
\end{align}
under the change of basis $\ket{n}\rightarrow U\ket{n}$, which follows from
$
\braket{m|U_{ST}K|n}
\rightarrow
\braket{m|U^{\dagger}U_{ST}KU|n}
=\braket{m|\left(U^{\dagger}U_{ST}U^*\right)K|n}.
$
Especially for orthogonal group elements, i.e., $U^TU=1$,
\begin{align}
\exp(iM)
&\rightarrow U^{\dagger}\exp(iM)U^*\notag\\
&=U^{T}\exp(iM)U\notag\\
&=\exp(iU^{T}MU).
\end{align}
Since $M$ is a real symmetric matrix, it can be diagonalized by an orthogonal transformation of basis.
In general, $U_{ST}={\rm diag}\left(e^{i\lambda_1},...,e^{i\lambda_{N}}\right)$ for $N$ bands after the change of basis.
The phases $\lambda_{1},...,\lambda_{N}$ can be erased by a phase rotation $U={\rm diag}\left(e^{-i\lambda_1/2},...,e^{-i\lambda_{N}/2}\right)$ so that we have $U_{ST}=1$.
As $I_{ST}^2=1$ holds for both spinless and spinful systems, the conclusion of this section does not depend on the existence of spin degrees of freedom.

\section{Berry Connection, Berry Curvature, and Sewing Matrices}

\subsection{Berry Connection and Berry Curvature}

The $2N\times 2N$ non-abelian Berry connection ${\cal A}$ and Berry curvature ${\cal F}$ are defined by
\begin{align}
{\cal A}_{nm}({\bf k})=i\braket{u_{n\bf k}|\nabla_{\bf k}|u_{m\bf k}},
\end{align}
and
\begin{align}
{\cal F}_{nm}({\bf k})=\nabla_{\bf k} \times {\cal A}_{nm}({\bf k})+i\left({\cal A}({\bf k})\times {\cal A}({\bf k})\right)_{nm},
\end{align}
where $n,m=1,...,2N$ are the band indices. 
The matrix elements ${\cal A}_{nm}$ and ${\cal A}_{nm}$ are considered three dimensional vectors in this definition of Berry curvature while the momentum ${\bf k}$ lives in a two-dimensional space.

The $U(1)$ part of the non-abelian Berry connection, i.e., the abelian Berry connection which is just called the ``Berry connection'' in many cases, is written in italic letter.
\begin{align}
A({\bf k})\equiv \sum_n{\cal A}_{nn}({\bf k})=\sum_ni\braket{u_{n\bf k}|\nabla_{\bf k}|u_{n\bf k}},
\end{align}
and
\begin{align}
F({\bf k})\equiv \sum_n{\cal F}_{nn}({\bf k})=\nabla_{\bf k} \times A({\bf k}).
\end{align}

Consider a gauge transformation
\begin{align}
\ket{u_{n\bf k}}\rightarrow \ket{v_{n\bf k}}=U({\bf k})\ket{u_{n\bf k}}=U_{mn}({\bf k})\ket{u_{m\bf k}},
\end{align}
where $U_{mn}({\bf k})=\braket{u_{m\bf k}|U({\bf k})|u_{n\bf k}}$.
The non-abelian Berry connection and curvature transform under the gauge transformation as
\begin{align}
{\cal A}_{nm}({\bf k})
\rightarrow (U^{\dagger}{\cal A}U)_{nm}({\bf k})+(U^{\dagger}i\nabla_{\bf k}U)_{nm}({\bf k})
\end{align}
and
\begin{align}
{\cal F}_{nm}({\bf k})
\rightarrow \left(U^{\dagger}({\bf k}){\cal F}({\bf k})U({\bf k})\right)_{nm},
\end{align}
which follows from their definition.

The transformation of $U(1)$ part is then found after taking trace over the above transformations.
\begin{align}
A({\bf k})
\rightarrow  A({\bf k})+i\nabla_{\bf k}\log\det U({\bf k})
\end{align}
and
\begin{align}
F({\bf k})
\rightarrow  F({\bf k}).
\end{align}

\subsection{Berry Connection under $C_{2z}$, $T$, and $C_{2z}*T$ Symmetries}

Consider the case where both $C_{2z}$ and $T$ symmetry is present. Sewing matrices $B$ and $D$ are defined as
\begin{align}
T\ket{u_{n\bf k}}
&=B_{mn}({\bf k})\ket{u_{m\bf -k}},\notag\\
C_{2z}\ket{u_{n\bf k}}
&=D_{mn}({\bf k})\ket{u_{m\bf -k}}.
\end{align}
They are unitary matrices because $T$ and $D$ are (anti)unitary operators which preserves the norm of wavefunctions.
We see explicitly that
\begin{align}
\delta_{nm}
&=\braket{u_{n\bf k}|u_{m\bf k}}\notag\\
&=\braket{u_{n\bf k}|u_{m\bf k}}^*\notag\\
&=\braket{Tu_{n\bf k}|Tu_{m\bf k}}\notag\\
&=\sum_{p,q}B^*_{pn}({\bf k})B_{qm}({\bf k})\braket{u_{p\bf -k}|u_{q\bf -k}}\notag\\
&=\sum_{p}B^*_{pn}({\bf k})B_{pm}({\bf k})
\end{align}
and
\begin{align}
\delta_{nm}
&=\braket{u_{n\bf k}|u_{m\bf k}}\notag\\
&=\braket{C_{2z}u_{n\bf k}|C_{2z}u_{m\bf k}}\notag\\
&=\sum_{p,q}D^*_{pn}({\bf k})D_{qm}({\bf k})\braket{u_{p\bf -k}|u_{q\bf -k}}\notag\\
&=\sum_{p}D^*_{pn}({\bf k})D_{pm}({\bf k}).
\end{align}

Using $T^2=\left(C_{2z}\right)^2=-1$ and unitarity of $B$ and $D$ matrices,
\begin{align}
T\ket{u_{n\bf -k}}
&=-B_{nm}({\bf k})\ket{u_{m\bf k}},\notag\\
C_{2z}\ket{u_{n\bf -k}}
&=-D^*_{nm}({\bf k})\ket{u_{m\bf k}}.
\end{align}
Thus we have
\begin{align}
B_{mn}({\bf -k})
&=-B_{nm}({\bf k}),\notag\\
D_{mn}({\bf -k})
&=-D^*_{nm}({\bf k}).
\end{align}

$T$ and $C_{2z}$ relates the non-abelian Berry connections at time-reversal momenta.
\begin{align}
{\cal A}^*_{nm}({\bf k})
&=
(B^{\dagger}({\bf k}){\cal A}({\bf -k})B({\bf k}))_{nm}
-(B^{\dagger}({\bf k})i\nabla_{\bf k}B({\bf k}))_{nm},\notag\\
{\cal A}_{nm}({\bf k})
&=-(D^{\dagger}({\bf k}){\cal A}({\bf -k})D({\bf k}))_{nm}+(D^{\dagger}({\bf k})i\nabla_{\bf k}D({\bf k}))_{nm}.
\end{align}

The space-time inversion $I_{ST}$ is the combination of $C_{2z}$ and $T$.

\begin{align}
G({\bf k})_{pn}
&=\braket{u_{p\bf k}|C_{2z}*T|u_{n\bf k}}\notag\\
&=\sum_m\braket{u_{p\bf k}|C_{2z}|u_{m\bf -k}}\braket{u_{m\bf -k}|T|u_{n\bf k}}\notag\\
&=\sum_mD_{pm}({\bf -k})B_{mn}({\bf k}).
\end{align}
Unitarity and the following relations also follow from that of $B$ and $D$.
\begin{align}
G({\bf k})=G^T({\bf k}),\quad G({\bf -k})=B({\bf k})G^*({\bf k})B^T({\bf k}).
\end{align}
The above relations can be derived noting that $C_{2z}*T=T*C_{2z}$ so that $G({\bf k})_{pn}=\sum_mD_{pm}({\bf -k})B_{mn}({\bf k})=\sum_mB_{pm}({\bf -k})D_{mn}({\bf k})$.
$C_{2z}*T$ gives a constraint on the non-abelian Berry connection which is a combination of $T$ and $C_{2z}$ constraints.
\begin{align}
{\cal A}_{nm}({\bf k})
&=-(G^{\dagger}({\bf k}){\cal A}({\bf k})G({\bf k}))^*_{nm}+(G({\bf k})i\nabla_{\bf k}G^{\dagger}({\bf k}))_{nm}.
\end{align}

The $U(1)$ part of the Berry connection satisfies
\begin{align}
A({\bf k})
&=+A({\bf -k})-i\nabla_{\bf k} \log \det B({\bf k})\notag\\
A({\bf k})
&=-A({\bf -k})+i\nabla_{\bf k} \log \det D({\bf k})
\end{align}
and
\begin{align}
A({\bf k})=-\frac{i}{2}\nabla_{\bf k}\log \det G({\bf k}),
\end{align}
combining the above two. The Berry connection is locally a pure gauge, and thus the Berry curvature is trivial.

\subsection{Gauge transformation of the sewing matrices $B$, $D$, and $G$}

Consider a gauge transformation
\begin{align}
\ket{u_{n\bf k}}\rightarrow \ket{v_{n\bf k}}=U({\bf k})\ket{u_{n\bf k}}=U_{mn}({\bf k})\ket{u_{m\bf k}},
\end{align}
Left and right hand side of each of the equation
\begin{align}
T\ket{u_{n\bf k}}
&=B_{mn}({\bf k})\ket{u_{m\bf -k}},\notag\\
C_{2z}\ket{u_{n\bf k}}
&=D_{mn}({\bf k})\ket{u_{m\bf -k}}.
\end{align}
is then expressed in the new basis as
\begin{align}
T\ket{u_{n\bf k}}
&=TU^{\dagger}_{pn}\ket{v_{p\bf k}}
=U^{T}_{pn}({\bf k})T\ket{v_{p\bf k}},\notag\\
C_{2z}\ket{u_{n\bf k}}
&=C_{2z}U^{\dagger}_{pn}\ket{v_{p\bf k}}
=U^{\dagger}_{pn}({\bf k})C_{2z}\ket{v_{p\bf k}},\notag\\
B_{mn}({\bf k})\ket{u_{m\bf -k}}
&=B_{mn}({\bf k})U^{\dagger}_{qm}({\bf -k})\ket{v_{q\bf -k}},\notag\\
D_{mn}({\bf k})\ket{u_{m\bf -k}}
&=D_{mn}({\bf k})U^{\dagger}_{qm}({\bf -k})\ket{v_{q\bf -k}},
\end{align}
so that we have
\begin{align}
T\ket{v_{p\bf k}}
&=U^*_{np}({\bf k})B_{mn}({\bf k})U^{\dagger}_{qm}({\bf -k})\ket{v_{q\bf -k}}\notag\\
&=(U^{\dagger}({\bf -k})B({\bf k})U^*({\bf k}))_{qp}\ket{v_{q\bf -k}},\notag\\
C_{2z}\ket{v_{p\bf k}}
&=U_{np}({\bf k})D_{mn}({\bf k})U^{\dagger}_{qm}({\bf -k})\ket{v_{q\bf -k}}\notag\\
&=(U^{\dagger}({\bf -k})D({\bf k})U({\bf k}))_{qp}\ket{v_{q\bf -k}}.
\end{align}

We have found the transformation of $B$ and $D$ matrices.
\begin{align}
B({\bf k})&\rightarrow U^{\dagger}({\bf -k})B({\bf k})U^*({\bf k}),\notag\\
D({\bf k})&\rightarrow U^{\dagger}({\bf -k})D({\bf k})U({\bf k}).
\end{align}
From this we find the gauge transformation of $G$ matrix.
\begin{align}
G({\bf k})&\rightarrow U^{\dagger}({\bf k})G({\bf k})U^*({\bf k}).
\end{align}

\section{Proof of $Z_2$ change by a partner exchange of Weyl pairs}

\subsection{$2P_{\rm I}$ as a Topological Invariant of the 1D Subsystem with $T$ and $C_{2z}$ Symmetries}

We show here that $2P^{\rm I}$ is a topological invariant of a 1D subsystem with $T$ and $C_{2z}$ symmetries where $T^2=(C_{2z})^2=-1$. 
First, we show it is quantized to an integer.
In a time-reversal invariant one-dimensional system, we can define the time-reversal polarization
\begin{align}
P_T=P^{\rm I}-P^{\rm II}=2P^{\rm I}-P_{\rho}\in {\bb Z},
\end{align}
which is gauge dependent, but quantized to an integer.\cite{Fu-Kane}
In the presence of $C_{2z}$ symmetry, the charge polarization $P_{\rho}$ is also quantized.

\begin{align}
P_{\rho}
&=\frac{1}{2\pi}\int^{\pi}_{-\pi} d{\bf k} \cdot A({\bf k})\notag\\
&=\frac{1}{2\pi}\int^{\pi}_0 d{\bf k}\cdot A({\bf k})+\frac{1}{2\pi}\int^{0}_{-\pi} d{\bf k}\cdot A({\bf k})\notag\\
&=\frac{1}{2\pi}\int^{\pi}_0 d{\bf k}\cdot A({\bf k})+\frac{1}{2\pi}\int^{\pi}_0 d{\bf k}\cdot A({\bf -k})\notag\\
&=\frac{i}{2\pi}\int^{\pi}_0\nabla_{\bf k} \log \det D({\bf k})\notag\\
&=\frac{i}{2\pi} \log \frac{\det D({\pi})}{\det D({0})}\in {\bb Z},
\end{align}
because $\det D(\Gamma_i)=1$ at any TRIM $\Gamma_i$ due to Kramers pairs having eigenvalue pairs $i$ and $-i$ under $D$.

Thus $2P^{\rm I}$ is quantized to an integer.
\begin{align}
2P^{\rm I}=P_T+P_{\rho}\in {\bb Z}.
\end{align}

Now we show the gauge invariance of $2P^{\rm I}$.
Under the $U(1)$ gauge transformation $\ket{u^{\rm s}_{\alpha\bf k}}\rightarrow \ket{u^{\rm s}_{\alpha\bf k}}= e^{i\theta^{\rm s}_{\alpha\bf k}}\ket{u^{\rm s}_{\alpha\bf k}}$, partial polarization $P^{\rm I}$ transforms as
\begin{align}
P^{\rm I}
&=\frac{1}{2\pi}\oint d{\bf k} \cdot{\rm Tr}_{\rm (I)}{\cal A}({\bf k})\notag\\
&\rightarrow \frac{1}{2\pi}\oint d{\bf k} \cdot{\rm Tr}_{\rm (I)}{\cal A}({\bf k})-\oint \frac{d\theta^I}{2\pi}\notag\\
&=P^{\rm I}-\oint \frac{d\theta^I}{2\pi},
\end{align}
where ${\rm Tr}_{\rm (I)}$ is a partial trace over $\ket{u^{\rm I}_{\alpha \bf k}}$ for all $\alpha$, $\theta^{\rm I}=\sum_{\alpha}\theta_{\alpha}^{\rm I}$, and the winding number $\oint \frac{d\theta^{\rm I}}{2\pi}$ is an arbitrary integer. 
Thus $2P_{\rm I}$ is $U(1)$ invariant mod 2.
Invariance under non-abelian gauge transformation can be seen from the expression
\begin{align}
P^{\rm I}
&=\frac{1}{2\pi}\left[\int^{\pi}_0 d{\bf k}\cdot A({\bf k})+i\log\frac{\Pf B({\pi})}{\Pf B({0})}\right]
\end{align}
presented by Fu and Kane\cite{Fu-Kane}. 
The abelian Berry connection is manifestly invariant.
$A({\bf k})\rightarrow A({\bf k})+{\rm Tr}U^{\dagger}({\bf k})i\nabla U({\bf k})= A({\bf k})+i\nabla\log \det U({\bf k})= A({\bf k})$ for $\det U({\bf k})=1$.
The matrix $B$ transformation as $B({\bf k})\rightarrow U^{\dagger}({\bf -k})B({\bf k})U^*({\bf k})$ under $\ket{u_{n\bf k}}\rightarrow U_{mn}({\bf k})\ket{u_{m\bf k}}$.
Using the identity $\Pf (X^TBX)=\det (X) \Pf (B)$, the Pfaffian part is also invariant.

This invariant $2P^{\rm I}$ was first noted by Lau, van den Brink, and Ortix \cite{mirror_insulator} as a topological invariant of the topological mirror insulator where $T^2=M^2=-1$. Notice that the two-fold rotation $C_{2z}$ reduces to a mirror operation on any time-reversal invariant 1D subsystem.

\subsection{Strategy of the Proof}

We calculate the change of $Z_2$ invariant due to movements of Weyl points by using the formula
\begin{align}
\Delta
&=P_T(\pi)-P_T(0)\notag\\
&=2P^{\rm I}(\pi)-2P^{\rm I}(0)-\left(P_{\rho}(\pi)-P_{\rho}(0)\right),
\end{align}
where $P_T(0)$ ($P_T(\pi)$) is the time-reversal polarization of a TRI 1D system passing through the two TRIM points with $k_x=0$ ($k_x=\pi$).
In an insulating phase, we can have a continuous gauge as Chern number in the whole BZ is zero.
In the gauge, we have
\begin{align}
P_{\rho}(\pi)-P_{\rho}(0)=\frac{1}{2\pi}\int_{\rm EBZ}dk_xdk_y F_z({\bf k})=0,
\end{align}
where $F=0$ from the spacetime inversion symmetry.
EBZ is the effective Brillouin zone of a time-reversal system which is a half Brillouin zone with TRI 1D systems at $k_x=0$ and $k_x=\pi$ as its boundary.
Thus the $Z_2$ invariant is expressed as a partial polarization pump
\begin{align}
\Delta=2P^{\rm I}(\pi)-2P^{\rm I}(0)
\end{align}
in a continuous gauge when the spacetime inversion symmetry is present. 
The two times partial polarization will change only if the TRI 1D system closes a gap since it is a topological invariant.
From now on, we will prove that when two Weyl point whose position in BZ is related by time-reversal cross the TRI 1D system, $2P^{\rm I}$ changes by one so that partner exchange of Weyl points leads to a change of $Z_2$ invariant.
Such crossing occurs even number of times in a partner preserving process so that the $Z_2$ invariant does not change. [See Fig.~\ref{partial}(a,b).]

\subsection{Change of the partial polarization at the gap closing}

\begin{figure*}[h!]
  \centering
    \includegraphics[width=\textwidth]{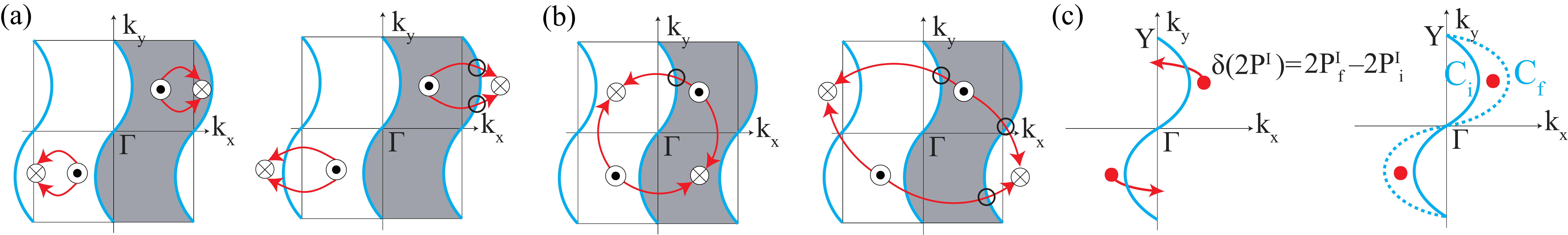}
  \caption{
Trajectory of gapless points in the 2D Weyl semimetal phase.
(a,b): 
Arrows indicate the trajectories which the Weyl points follow in the process of phase transition.
A blue solid line indicates a time-reversal invariant 1D subsystem.
$\odot$ and $\otimes$ are locations for pair-creation and pair-annihilation.  
In the partner preserving process shown in (a), Weyl points related by
time reversal symmetry cross the blue line even number of times.
In the partner exchanging process shown in (b), Weyl points related by
time reversal symmetry cross the blue line odd number of times.
(c): 
Red dots are Weyl points with Berry phase $\pm \pi$. 
The solid curve ($C_i$) is the originally defined one-dimensional time-reversal invariant system. 
The dotted curve $C_f$ can be continuously deformed to $C_i$ without a gap-closing after the Weyl points pass through $C_i$.
The change of topological invariant of $C_i$ due to the crossing of Weyl points can thus be calculated as the difference of topological invariant between the curve $C_i$ and $C_f$.
}
\label{partial}
\end{figure*}

We take a time-reversal invariant 1D systems avoiding pair creation and annihilation as in  Fig.~\ref{partial}.
The gauge is chosen so that the wavefunction is discontinuous only at the singularities.
It is possible because there is no Berry curvature punching out singularities. 
In this gauge, we can use the expression
\begin{align}
P^{\rm I}
&=\frac{1}{2\pi}\left[\int^{\pi}_0 d{\bf k}\cdot A({\bf k})+i\log\frac{\Pf B({\pi})}{\Pf B({0})}\right].
\end{align}

The time-reversal invariant curve $C_f$ can be deformed adiabatically to the $C_i$ line after Weyl points pass through. [See Fig.~\ref{partial}(c).]
Thus we calculate the change of (two times) partial polarization by the crossing of Weyl points as the difference of it between the curve $C_i$ and $C_f$.

\begin{align}
\delta\left(2P^{\rm I}\right)
&=\delta\left(\frac{1}{\pi}\int^{\pi}_0 d{\bf k}\cdot A({\bf k})-2\times\frac{1}{2\pi i}\log\frac{\Pf B({\pi})}{\Pf B({0})}\right)\notag\\
&=\frac{1}{\pi}\left[\int_{C_f/2} d{\bf k}\cdot A({\bf k})-\int_{C_i/2} d{\bf k}\cdot A({\bf k})\right]\notag\\
&\quad-2\times\frac{1}{2\pi i}\left[\log\frac{\Pf B_f({\pi})}{\Pf B_f({0})}-\log\frac{\Pf B_i({\pi})}{\Pf B_i({0})}\right]\notag\\
&=\frac{1}{\pi}\oint_{C} d{\bf k}\cdot A({\bf k})\notag\\
&\quad-2\times\frac{1}{2\pi i}\log\left[\frac{\Pf B_f({\pi})}{\Pf B_i({\pi})}\left(\frac{\Pf B_f({0})}{\Pf B_i({0})}\right)^{-1}\right]\notag\\
&=\frac{1}{\pi}\oint_{C} d{\bf k}\cdot A({\bf k})\quad {\rm (mod \;2)}\notag\\
&=1\quad {\rm (mod \;2)},
\end{align}
where $C_{i,f}/2$ is the $k_y>0$ part of the curve $C_{i,f}$, we used the single-valuedness of the sewing matrix $B$ at $\Gamma$ and Y points in the fourth line, and used the property of Weyl points at the last line.
We have now proven that the $Z_2$ invariant changes through a partner exchange.

\section{Alternative Proof}

\subsection{Alternative Proof by 3D Embedding}

\begin{figure*}[h!]
  \centering
    \includegraphics[width=\textwidth]{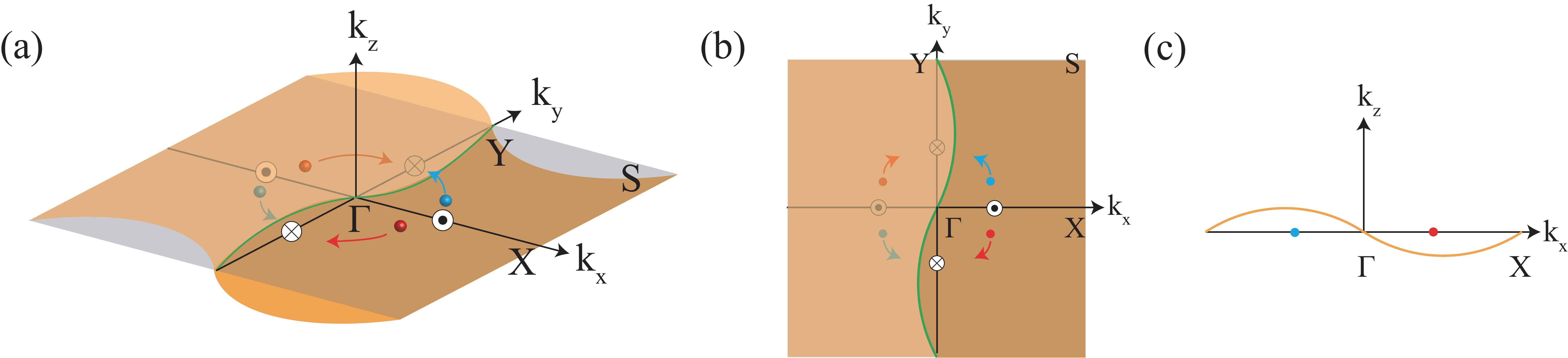}
  \caption{Schematic picture describing pair-creation/pair-annihilation process. 
(a) Bird's eye view. 
The orange sheet is a deformed two-dimensional system. 
The green line indicates the intersection between the deformed 2D system and $k_z=0$ plane. 
Blue and red spheres are monopoles and antimonopoles.
$\odot$ ($\otimes$) marks the location where pair-creation (pair-annihilation) happens.
(b) Top view.
(c) Front view.}
\label{deform}
\end{figure*}

Here we present an alternative proof of the change of $Z_2$ invariant due to the pair change of Weyl points.
This proof starts by embedding the 2D BZ of our system into a 3D BZ of a system with the same symmetry, $C_{2z}$ and $T$.
After a suitable embedding, the proof goes by following the analysis of Murakami et al. \cite{Murakami}.
Let $k_z=0$ plane of the 3D BZ be the 2D BZ of our system.
The 2D Weyl points will be lifted into 3D Weyl points rather than a line node in general because there is no protection of $\pi$ Berry flux out of the $C_{2z}*T$ invariant plane, i.e., out of the $k_z=0$ or $k_z=\pi$ plane.
In fact, we can choose an embedding to get 3D Weyl points.
Next, consider the deformed time-reversal invariant two-dimensional system illustrated as the orange sheet in Fig.~\ref{deform}.
In this construction, a Weyl point passes through the EBZ of the orange sheet.
This change the $Z_2$ invariant of the orange sheet as claimed by Murakami et al.\cite{Murakami} and it will be proved explicitly below.
The sheet can be adiabatically deformed into $k_z=0$ before/after the creation/annihilation of Weyl points.
Thus the change of $Z_2$ invariant of the 2D system at $k_z=0$ is the same as the $Z_2$ change of the orange sheet, and the proof ends.

\subsection{Explicit Proof of the Claim by Murakami and Kuga}

For completeness, we prove the statement in \prb {\bf 78}, 165313.
In the paper, the authors claimed that the $Z_2$ invariant of a 2D TRI BZ changes when a Weyl point pass through its EBZ, noting that the Berry flux of the half-BZ determines $Z_2$ invariant (up to some extra term) as proved by Moore and Balents \cite{Moore-Balents}.
The extra term is the Berry flux of the extended part which is added to make EBZ closed, i.e., homotopically equivalent to a sphere.
It is not obvious why the extra term does not contribute to the $Z_2$ change without having analytic expression of it.
An explicit proof has not been presented anywhere while the statement is believed to be valid. 
On the other hand, Fu and Kane argued in the appendix of \prb {\bf 74}, 195312 that the extra term corresponds to the pump of $2P^{\rm I}$.
In this subsection, we show using the expression of the 2D $Z_2$ invariant as a time-reversal polarization pump that the invariant changes when a 3D Weyl points pass through it on a EBZ. 
This proof will clarify why is it possible to calculate the change of 2D $Z_2$ invariant from the Berry flux jump within a EBZ.

\begin{figure*}[h!]
  \centering
    \includegraphics[width=\textwidth]{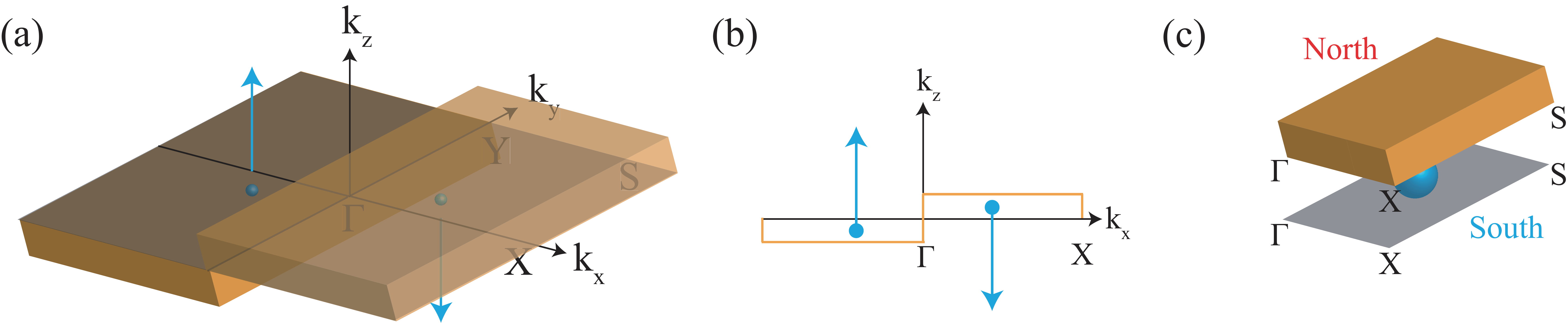}
  \caption{(a) The geometry TRI 2D surfaces and the position of Weyl points. 
(a) Bird's eye view. 
The orange sheet is a deformed two-dimensional time-reversal invariant system. 
(b) Front view.
(c) A Weyl point is surrounded by two sheets, each of which represent a 2D patch where wave function is smoothly defined.}
\label{Murakami}
\end{figure*}

Consider the the Brillouin zone of a time-reversal invariant system as in Fig.~\ref{Murakami}-(a,b).
$C_{2z}$ symmetry is not assumed.
The Blue balls are Weyl points with positive chirality.
The orange sheet is a deformed 2D time-reversal invariant sub-BZ sharing the same EBZ boundary with the $k_z=0$ plane.
The $k_z=0,k_y=\pm \pi$ lines are also common.
Weyl points did not pass through the $k_z=0$ plane yet. 
After Weyl points cross, the orange sheet can be adiabatically deformed to $k_z=0$ plane. 
Thus the difference of the $Z_2$ invariant between the orange sheet and $k_z=0$ plane gives the change of $Z_2$ invariant under the crossing procedure.

Let us choose a continuous gauge for each 2D BZ. 
It is always possible because they are gapped time-reversal invariant 2D BZ with vanishing Chern number. 
Then the time-reversal polarization is well defined, and the $Z_2$ invariant is expressed as

\begin{align}
\Delta=2P^{\rm I}(\pi)-2P^{\rm I}(0)-\frac{1}{2\pi}\int_{\rm EBZ}d{\bf S}\cdot F({\bf k}).
\end{align}
$d{\bf S}=d^2k\hat{n}$ where $\hat{n}$ is a unit vector normal to the 2D surface $dS$, and we used $P_{\rho}(\pi)-P_{\rho}(0)=\frac{1}{2\pi}\int_{\rm EBZ}F$ for a continuous gauge.
While it is possible to choose a continuous gauge in each 2D BZ, there should be a discrete transition between them since Weyl points are enclosed by the sheets.
Let us call the gauge in the orange sheet and $k_z=0$ plane as north and south, respectively (Fig.~\ref{Murakami}-(c)). The difference of $Z_2$ invariant is

\begin{align}
\delta\Delta
&=\Delta_N-\Delta_S\notag\\
&=2P_N^{\rm I}(\pi)-2P_N^{\rm I}(0)-\left(2P_S^{\rm I}(\pi)-2P_S^{\rm I}(0)\right)\notag\\
&\quad-\frac{1}{2\pi}\left(\int_{N,\rm EBZ}d{\bf S}\cdot F({\bf k})-\int_{S,\rm EBZ}d{\bf S}\cdot F({\bf k})\right)\notag\\
&=-\frac{1}{2\pi}\left(\int_{N,\rm EBZ}d{\bf S}\cdot F({\bf k})-\int_{S,\rm EBZ}d{\bf S}\cdot F({\bf k})\right)\quad ({\rm mod \; 2})\notag\\
&=-\frac{1}{2\pi}\oint_{\rm Weyl}d{\bf S}\cdot F({\bf k}) \quad ({\rm mod \; 2})\notag\\
&=-1 \quad ({\rm mod \; 2}),
\end{align}
where we used the mod 2 invariant of $2P^{\rm I}$ under the discrete gauge transformation, i.e., the transition function, in the second line, and used the definition of the chirality of Weyl points in the last line.

\section{Application: Kane-Mele Model}

\subsection{Tight-Binding Model}

\begin{figure*}[h!]
\centering
\includegraphics[width=\textwidth]{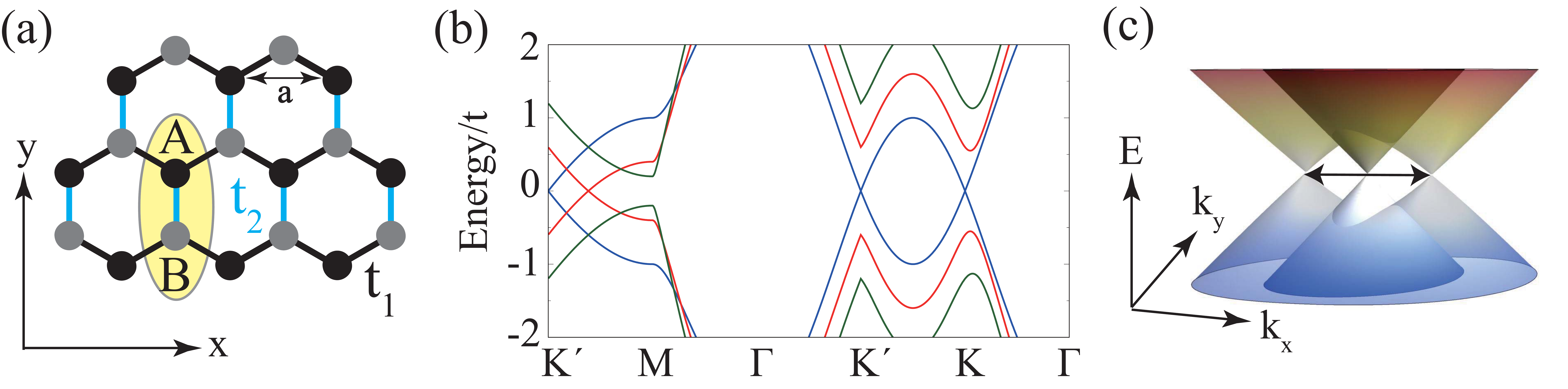}
\caption{
(a) Kane-Mele model on the honeycomb lattice strained along the $y$-direction (armchair direction).
The unit cell is composed of two sublattice sites A and B.
Due to the strain, the hopping amplitude $t_2$ between horizontal zigzag wires is different from the hopping $t_1$ within a zigzag wire.
The strained system is still symmetric under the $C_{2z}$ whose axis penetrates the center of a bond with hopping amplitude $t_{2}$.
(b) The band structure of the strained Kane-Mele model along the high symmetric lines.
Spin-orbit coupling is not included here,
Blue, red, and green line are plotted for $r=1$, $r=1.6$, and $r=2.2$.
Here $\Gamma=(0,0)$, $M=(0,\frac{2\pi}{\sqrt{3}a})$, $K=(\frac{4\pi}{3a},0)$, and $K'=(\frac{2\pi}{3a},\frac{2\pi}{\sqrt{3}a})$ are high-symmetric points.
As $r$ increases, the Dirac points at $K$ and $K'$ moves toward $M$ and the band gap opens via a pair-annihilation of Dirac points at $M$.
(c) The band structure in the presence of Rashba spin-orbit coupling near a Dirac point at $r=1.6$.
The Dirac point with four-fold degeneracy splits into two Weyl points with two-fold degeneracy.
}
\label{fig:graphene_tight-binding}
\end{figure*}

To demonstrate the unconventional topological phase transition, we construct
a simple model Hamiltonian, a modified Kane-Mele model on the honeycomb lattice.
Consider
\begin{align}
H=\sum_{\braket{ij}}t_{ij}c^{\dagger}_{i}c_{j}+\sum_{\braket{\braket{ij}}}i\lambda_{SO}c^{\dagger}_i\bm{s}\cdot\hat{\bm{\nu}}_{ij}c_j+H_{R},
\end{align} 
where $t_{ij}=t$ is the nearest neighbor hopping amplitude and 
$\lambda_{SO}$ is the spin-dependent hopping between next nearest neighbor sites.
These two terms define the usual Kane-Mele Hamiltonian describing a QSHI.
Here $\hat{\bm{\nu}}_{ij}=(\bm{d}^{1}_{ij}\times \bm{d}^{2}_{ij})/|(\bm{d}^{1}_{ij}\times \bm{d}^{2}_{ij})|$ where $\bm{d}^{1}_{ij}$ and $\bm{d}^{2}_{ij}$
are bond vectors between nearest-neighbor sites
on which the electron traverses when hopping from $j$ to $i$.
Additionally, we include the nearest-neighbor Rashba term 
\begin{align}
H_R=i\lambda_R\sum_{\braket{ij}}c^{\dagger}_i\hat{z}\cdot({\bf s}\times\hat{\bf d}_{ij})c_j
\end{align}
with $\hat{\bm{d}_{ij}}=\bm{d}_{ij}/|\bm{d}_{ij}|$ 
induced by the substrate or 
electric field breaking inversion symmetry.
As shown in Ref.~\onlinecite{Kane-Mele},
QSHI persists when $\lambda_{R}<2\sqrt{3}\lambda_{SO}$
whereas a semimetal with quadratic band crossing appears when $\lambda_{R}>2\sqrt{3}\lambda_{SO}$. Thus one can achieve only an insulator-semimetal transition by varying
the ratio $\lambda_{R}/\lambda_{SO}$.
In fact, when $\lambda_{R}>2\sqrt{3}\lambda_{SO}$, the band degeneracy of the semimetal
phase is protected by $C_{3v}$ symmetry of the honeycomb lattice.
Therefore to achieve the transition from a QSHI to a normal band insulator, 
it is necessary to add additional terms breaking $C_{3v}$ symmetry.

To achieve a topological phase transition, we introduce hopping anisotropy between nearest neighbor sites, which can be induced by applying uniaxial strain~\cite{Hasegawa,YWSon}.
Explicitly, as shown in Fig.~\ref{fig:graphene_tight-binding} (a), we choose $t_{ij}=t_{2}$ for bonds parallel to the $y$-axis whereas $t_{ij}=t_{1}$ for the other two bonds along horizontal zig-zag chains.
When $t_{1}\neq t_{2}$, the system has only a two-fold rotation symmetry ($C_{2z}$) about a $z$-axis passing the mid-point of the bond with hopping amplitude $t_{2}$.
We examine the evolution of the band structure varying $r\equiv t_{2}/t_{1}$ which is tunable by controlling strain. 
$\lambda_{R}$ and $\lambda_{SO}$ are fixed but are chosen to satisfy 
$\lambda_{R}<2\sqrt{3}\lambda_{SO}$, which ensures that the system is a QSHI when $t_{2}=t_{1}$.

\subsection{Low Energy Analysis Near $M$ Point}

Changing the ratio $r\equiv t_2/t_{1}$, one can observe a gap-closing near the $M$ point.
The effective Hamiltonian near the $M$ point is given by
\begin{align}
H(\bm{q})
&=(t_2-2t_1+t_1q^2)\sigma_{x}-2t_1q_y \sigma_{y}\notag\\
&\quad
+8 \lambda_{SO}q_x\sigma_{z}s_{z}
+2\lambda_{R}\sigma_{y}s_{x}
\end{align}
where $(q_x,q_y)=(\frac{k_xa}{2},\frac{\sqrt{3}k_ya}{2})-(0,\pi)$, $q=\sqrt{q_x^2+q_y^2}$, and $\sigma_{x,y,z}$ ($s_{x,y,z}$) are Pauli matrices for sublattice (spin) degrees of freedom.
The energy eigenvalue of $H(\bm{q})$ is given by
\begin{align}
E^{(\pm)}_{\pm}({\bf q})
&=(\pm)\sqrt{\left(f({\bf q})\right)^2+\left(g_{\pm}({\bf q})\right)^2},\notag\\
f({\bf q})
&=t_2-2t_1+t_1q^2,\notag\\
g_{\pm}({\bf q})
&=\sqrt{(2t_1q_y)^2+(8 \lambda_{SO}q_x)^2}\pm 2\lambda_R.
\end{align}
The band gap closes when 
\begin{align}
q_x^2+q_y^2
=2-r,\quad
\frac{q_x^2}{(\lambda_R/t_1)^2}+\frac{q_y^2}{(\lambda_R/4\lambda_{SO})^2}
= 1
\end{align}
 are satisfied. The solutions of
the coupled equations
describe a circle and an ellipse, respectively, both centered at $\bm{q}=0$ in the $(q_{x},q_{y})$ plane.
Explicitly, one can find four gap-closing points, each of which describes a 2D Weyl point, when $r_{c1}<r<r_{c2}$ where $r_{c1}=2-\left(\lambda_R/4\lambda_{SO}\right)^2$ and $r_{c2}=2-\left(\lambda_R/t_1\right)^2$ where
we assume $t_{1}>4\lambda_{SO}$.
At the critical point with $r_{c1}$ ($r_{c2}$), the band gap closes at two points $(q_x,q_y)=(\pm \lambda_R/4\lambda_{SO},0)$ ($(q_x,q_y)=(0,\pm\lambda_R/t_1$)) where
each gap-closing point describes an anisotropic Weyl point.
Thus, as $r$ increases between two critical points, four Weyl points are created on the $q_y=0$ line, and then they move along an elliptical trajectory until they are pair-annihilated on the $q_x=0$ line.

\subsection{Numerical Calculation}

\begin{figure*}[h!]
\centering
\includegraphics[width=\textwidth]{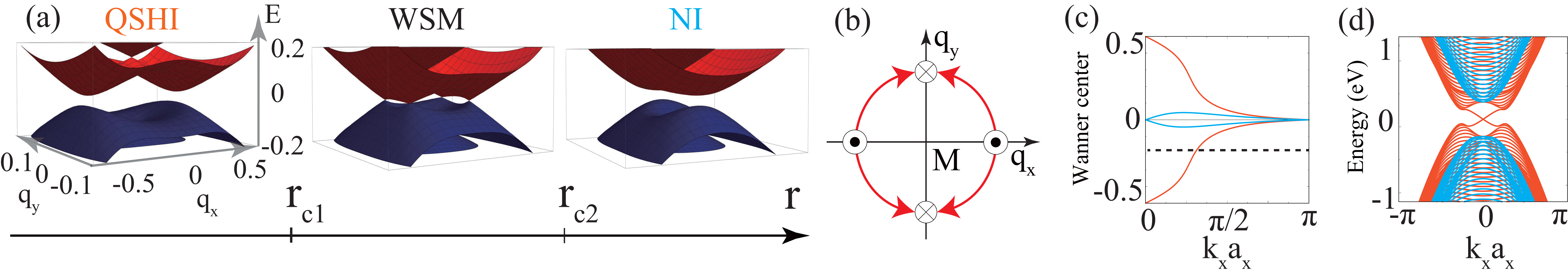}
\caption{
(a) Evolution of the band structure across a TPT in strained Kane-Mele model.
The spin-orbit coupling constants are fixed to be $\lambda_{SO}=\lambda_R=0.05t$.
The low energy band dispersion near the $M$ point is shown where
$q_x=k_xa/2$ and $q_y=\sqrt{3}k_ya/2-\pi$.
The bands are plotted at $r=1.90$, 1.96, and 2.02, respectively.
(b) Trajectory of WPs as $r$ increases in the Weyl semimetal phase. 
$\odot$ and $\otimes$ are the locations where pair-creation and pair-annihilation happen.
(c) Evolution of Wannier function centers. 
The orange line crosses the reference line (dashed) once, indicating the $Z_2$ nontrivial nature.
(d) Finite-size band structure.
The topological nontriviality is also confirmed by the presence of helical edge states. 
In both (c) and (d), orange and blue lines are plotted for $r=1.8$ and $r=2.2$, respectively.
}
\label{fig:graphene_phase_transition}
\end{figure*}

The band structure near the $M$ point is described in Fig.~\ref{fig:graphene_phase_transition} (a).
To prove that the 2D Weyl semimetal phase mediates a TPT,
we should compare the $Z_{2}$ invariant $\Delta$ of two insulating phases
existing when $r<r_{c1}$
and when $r>r_{c2}$, respectively.
For this purpose, we first compute the energy spectrum of a strip structure
of the honeycomb lattice having a finite size along
one direction while keeping the translational invariance along the other direction.
As shown in Fig.~\ref{fig:graphene_phase_transition} (d), when $r<r_{c1}$, one can clearly observe helical edge states
localized on the sample boundary, which is absent when $r>r_{c2}$.
Thus the system is a QSHI (normal insulator) when $r<r_{c1}$ ($r>r_{c2}$).
For further confirmation, we directly compute $\Delta$ numerically.
Since inversion symmetry is broken, one cannot use parity eigenvalues to evaluate $\Delta$. Instead we determine $\Delta$ by computing the change of the time-reversal polarization between $k_{x}=0$ and $k_{x}=\pi$. 
Since the time-reversal polarization is given by the difference in the Wannier function centers of Kramers pairs, one can determine $\Delta$ by studying how Wannier function centers of Kramers pairs evolve between $k_{x}=0$ and $k_{x}=\pi$~\cite{Fu-Kane,wilson_loop}. 
As shown in Fig.~\ref{fig:graphene_phase_transition} (c),
when $r<r_{c1}$ ($r>r_{c2}$), one can see partner-switching (no partner-switching) of Wannier functions when $k_{x}$ changes from 0 to $\pi$.
Since the partner-switching (no partner-switching) between Wannier states indicates the change of
the time-reversal polarization by 1 (0), one obtains
$\Delta=1$ ($\Delta=0$) when $r<r_{c1}$ ($r>r_{c2}$).


\section{Application: Few-Layer Black Phosphorus}

Here we propose that the novel TPT mediated by
pair-annihilation/pair-creation of 2D Weyl points can be observed in
few-layer black phosphorus systems under vertical electric field.
A recent experiment~\cite{KSKim} has shown that the band-gap of this system can be controlled by doping potassium on the surface,
thus the insulator-semimetal transition can be realized.
As shown in Fig.~\ref{fig:tight-binding}(a), a black phosphorus
is composed of puckered honeycomb layers stacked vertically.
For illustration, we construct a $ k\cdot p$ theory and a tight-binding Hamiltonian relevant to
the bilayer phosphorene. Though we focus on the bilayer system, for convenience,
the same idea can be generalized to thicker layered black phosphorus systems. 
The monolayer case is an exception in which the band gap is not tunable by electric field.

\subsection{Low Energy Effective Hamiltonian near ${\bf k}=0$}

Before studying the tight-binding model, we construct a general $k\cdot p$ Hamiltonian at $\Gamma$ point from symmetry considerations.
The low energy Hamiltonian can help understand the physics involved here as the tight-binding model for few-layer black phosphorus is complicated.
The symmetry group for a stacked phosphorene is a nonsymmorphic symmetry group isomorphic to $D^7_{2h}(\# 53)$ ($D^7_{2h}(\# 57)$) for odd (even) number of layers. 
In the presence of an electric field along the $z$-direction, the symmetry operations changing the sign of $z$ are broken so that its point group reduce to \{$1,M_x,M_y,C_{2z}$\}. 
The translational part of nonsymmorphic group elements is neglected here because the representation of nonsymmorphic group at $\Gamma$ behaves like a representation of its symmorphic part . 
Only (glide) mirror symmetries will be considered because $C_{2z}=M_{x}M_{y}$.

Consider the most general two-level Hamiltonian of a spinless system.
\begin{align}
H_0(k_x,k_y)
&=g_0(k_x,k_y)\sigma_0+g_x(k_x,k_y)\sigma_x\notag\\
&\quad+g_y(k_x,k_y)\sigma_y+g_z(k_x,k_y)\sigma_z,
\end{align}
where $\sigma_0=1$, and $\sigma_{x,y,z}$ are Pauli matrices for orbital degrees of freedom.
The chemical potential $g_0$ will be neglected below as it is not important for gap closing.
In $M_x$ invariant system, $M_xH_0(k_x,k_y)M_x^{-1}=H_0(-k_x,k_y)$, where $M_x^2=1$.
Because $M_x$ and $H_0$ commute on the $M_x$-invariant line with $k_x=0$, we can choose a basis in which they are diagonalized simultaneously.
In such a basis $M_x=\sigma_0$ or $\sigma_z$. 
For $M_x=\sigma_0$, Hamiltonian is even in $k_x$.
However, few-layer black phosphorus has linearly dispersing Dirac points on the line $k_x=0$ when bands are inverted by applying an out-of-plane electric field\cite{Choi}, which we demonstrate using a tight-binding model below.
$M_x=\sigma_z$ is thus appropriate for our system.
$M_x$ symmetry conditions for the representation are 
\begin{align}
g_{x,y}(k_x,k_y)
&=-g_{x,y}(-k_x,k_y),\notag\\
g_z(k_x,k_y)
&=g_z(-k_x,k_y).
\end{align}
Mirror operator $M_y$ is represented by $\sigma_0$ or $\sigma_z$ since it should commute with $M_x$.
First, consider $M_y=\sigma_z$.
The symmetry constraint $M_yH_0(k_x,k_y)M_y^{-1}=H_0(k_x,-k_y)$ gives
$
g_{x,y}(k_x,k_y)=-g_{x,y}(k_x,-k_y)$ and $g_z(k_x,k_y)=g_z(k_x,-k_y).
$
Combining the constraints from $M_x$ and $M_y$ symmetries, we have
$
g_{x,y}(k_x,k_y)=g_{x,y}(-k_x,-k_y)$ and $g_z(k_x,k_y)=g_z(-k_x,-k_y).
$
It is consistent with time-reversal symmetry $TH_0(k_x,k_y)T^{-1}=H_0(-k_x,-k_y)$ and commutation relations $[T,M_x]=[T,M_y]=0$ when $T=K$ ($\sigma_zK$) and $g_y=0$ ($g_x=0$).
The Hamiltonian compatible with the given symmetry representation is
$
H_0(k_x,k_y)=Ak_xk_y\sigma_x+(M-B_1k_x^2-B_2k_y^2)\sigma_z
$
and
$
H_0(k_x,k_y)=Ak_xk_y\sigma_y+(M-B_1k_x^2-B_2k_y^2)\sigma_z
$
up to quadratic order in momenta when $T=K$ and $T=\sigma_zK$, respectively.
In both cases, band gap can close only at the $\Gamma$ point, which is different from our system.
Thus, we should take $M_y=\sigma_0$. 
$M_y$ symmetry then imposes
\begin{align}
g_{x,y,z}(k_x,k_y)=g_{x,y,z}(k_x,-k_y).
\end{align}
The combination of $M_x$ and $M_y$ symmetry constraints gives
\begin{align}
g_{x,y}(k_x,k_y)
&=-g_{x,y}(-k_x,-k_y),\notag\\
g_z(k_x,k_y)
&=g_z(-k_x,-k_y).
\end{align}
It is consistent with $TH_0(k_x,k_y)T^{-1}=H_0(-k_x,-k_y)$ and $[T,M_x]=[T,M_y]=0$ when $T=K$ and $g_x=0$.
We have determined the representation of the symmetry operators as
\begin{align}
M_x=\sigma_z, \quad M_y=\sigma_0, \quad T=K,
\end{align}
and the Hamiltonian constrained by symmetries above are
\begin{align}
H_0(k_x,k_y)=Ak_x\sigma_y+(M-B_1k_x^2-B_2k_y^2)\sigma_z.
\end{align}
up to quadratic order in momenta, and $A$, $M$, $B_1$, and $B_2$ are constants. 
Gap closes at $(k_x,k_y)=(0,\pm \sqrt{M/B_2})$ for positive $M/B_2$. 
$M$ can be shifted by applying uniform perpendicular electric field to drive the insulator-semimetal transition.

Now we include spin degrees of freedom. 
The spinless Hamiltonian $H_0$ trivially extends to a $4\times 4$ matrix, i.e., $H_0\rightarrow s_0\otimes H_0$ where $s_0$ is the $2\times 2$ identity matrix for spin degrees of freedom.
The most general spin-orbit coupled Hamiltonian is constructed by imposing the following symmetries
\begin{align}
M_x=is_x\sigma_z,\quad M_y=is_y,\quad T=is_yK,
\end{align}
where $s_{x,y,z}$ are spin Pauli matrices.
We first impose space-time inversion symmetry $I_{ST}H_0(k_x,k_y)I_{ST}^{-1}=H_0(k_x,k_y)$ for $I_{ST}=M_xM_yT=-is_xK$.
Seven matrices 
$
s_y\sigma_y
,
s_z\sigma_x
,
s_y\sigma_z
,
s_x\sigma_z
,
s_x
,
s_y$,
and
$s_x\sigma_y$
which are commuting with $I_{ST}$ are basic building blocks of the spin-orbit coupled Hamiltonian.
$M_x$ and $M_y$ mirror symmetries determine the parity of functions multiplied by the seven matrices.
The spin-orbit coupled part of Hamiltonian is then
\begin{align}
H_{\lambda}(k_x,k_y)
&=\lambda_1s_y\sigma_y
+
\lambda_2k_ys_z\sigma_x
+
\lambda_3k_xs_y\sigma_z\notag\\
&\quad
+
\lambda_4k_ys_x\sigma_z
+
\lambda_5k_y s_x
+
\lambda_6k_xs_y,
\end{align}
up to first order in momenta.
Since the spin-orbit coupling parameters are small in black phosphorus, the effect of the terms which are second order in momentum is not considered here, unlike the case of the spinless Hamiltonian.
The spin-orbit couplings can change the normal insulator-Dirac semimetal transition of the spinless system into the normal insulator-Weyl semimetal-quantum spin Hall insulator transition.
We first discuss how each term affects Dirac points.
Among the terms, $\lambda_2k_ys_z\sigma_x$ is the only term which anti-commutes with the spinless Hamiltonian $H_0$.
Thus it can open a band gap when a Dirac point is at $k_y=k_0\ne 0$ like a mass term.
Other terms either split a Dirac point into two symmetry-protected Weyl points or induce velocity anisotropy between two Weyl points consisting of a Dirac point, but they do not open a band gap.
$\lambda_1$, $\lambda_3$, and $\lambda_5$ term split a Dirac point in $k_x$, $k_y$, and energy-direction, respectively. $\lambda_4$ and $\lambda_6$ induces velocity anisotropy between two Weyl points sitting at the same momentum.
Different spin-orbit coupling terms compete with each other, and the dominant term determines the fate of Dirac points.
This consideration shows that the topological phase transition from a Weyl semimetal to a quantum spin Hall insulating phase occur when $\lambda_2k_ys_z\sigma_x$ term becomes dominant over other terms.
Below let us suppose that $\lambda_2$ is larger than other momentum dependent spin-orbit terms $\lambda_{3,4,5,6}$. 
Then the resulting Hamiltonian which captures the leading effect of spin-orbit couplings is given by

\begin{align}
H(k_x,k_y)
&=Ak_x\sigma_y+(M-B_1k_x^2-B_2k_y^2)\sigma_z\notag\\
&\quad+\lambda_1s_y\sigma_y+\lambda_2k_ys_z\sigma_x.
\end{align}
This Hamiltonian can be diagonalized analytically as
\begin{align}
E^{(\pm)}_{\pm}
&=
\pm \sqrt{\left(f(k_x,k_y)\right)^2+\left(g_{\pm}(k_x,k_y)\right)^2},\notag\\
f(k_x,k_y)
&=M-B_1k_x^2-B_2k_y^2,\notag\\
g_{\pm}(k_x,k_y)
&=\sqrt{(Ak_x)^2+(\lambda_2k_y)^2}(\pm) |\lambda_1|.
\end{align}
The band gap closes at momenta satisfying
\begin{align}
&B_1k_x^2+B_2k_y^2=M,\notag\\
&(Ak_x)^2+(\lambda_2k_y)^2=(\lambda_1)^2.
\end{align}
As $M$ is varied, four Weyl points can appear because the number of intersection points of two conical sections is four in general.
At critical values
\begin{align}
M_{c1}=B_1\left(\frac{\lambda_1}{A}\right)^2,\quad M_{c2}=B_2\left(\frac{\lambda_1}{\lambda_2}\right)^2,
\end{align}
we have two gap closing points with linear-quadratic dispersion (anisotropic Dirac points) at $(k_x,k_y)=(\pm \lambda_1/A,0)$ and $(k_x,k_y)=(\pm \lambda_1/\lambda_2,0)$, respectively.
When other spin-orbit couplings are included, the trajectory of Weyl points is deformed, but one can still observe a NI-WSM-QSHI transition as long as $\lambda_2>\lambda_{3,4,5,6}$.

\subsection{Tight-Binding Model}
\label{sec:BP-tight-binding}

\begin{figure*}[h!]
\centering
\includegraphics[width=\textwidth]{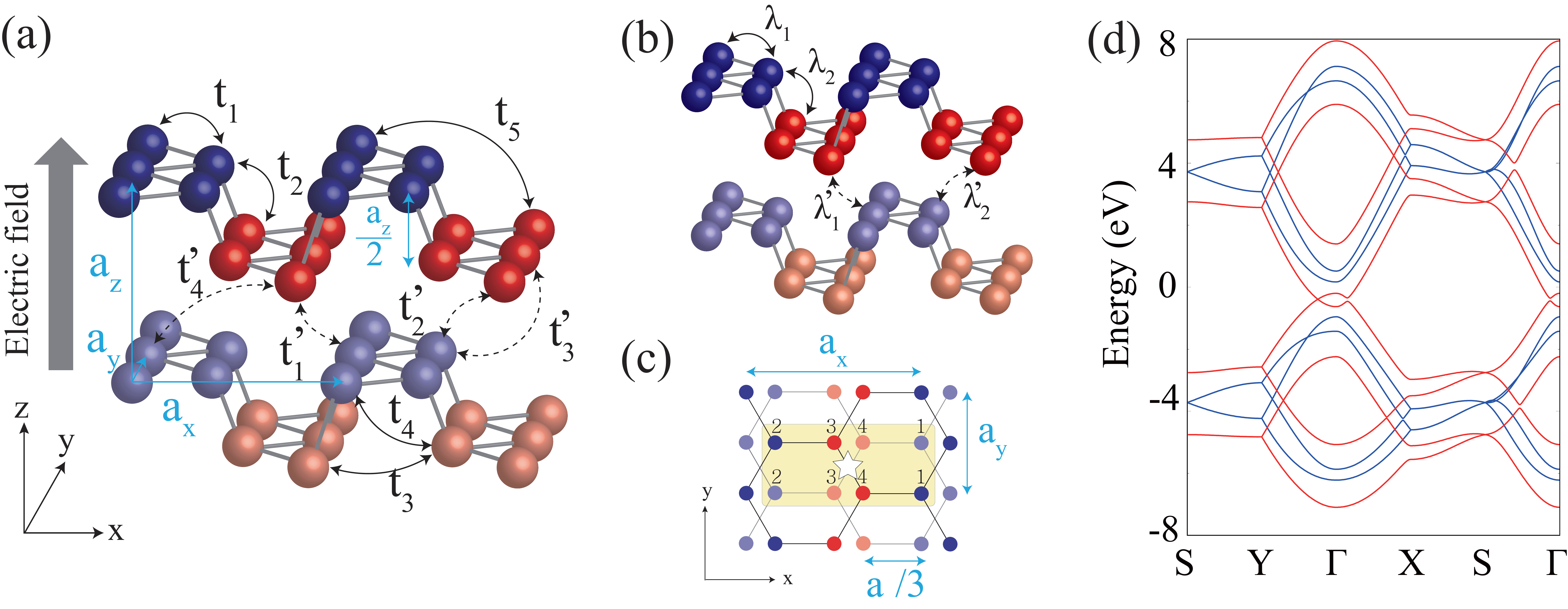}
\caption{
(a-c) Structure of a bilayer black phosphorus and hopping amplitudes. 
(a) Bird's-eye view.
Uniform static electric field $E_{\rm ext}$ is applied perpendicular to the bilayer sheet. Intralayer (interlayer) hoppings $t_{ij}$ ($t'_{ij}$) are shown. 
The height difference between two sublayers within a layer is taken to be $a_z/2$, thus all sublayers are equally spaced in the $z$-direction. $a_{x,y,z}$ indicates the unit lattice spacing in the $x,y,z$ direction.
Each sublayer is colored differently for clarity.
(b) Intralayer (interlayer) spin-orbit couplings $\lambda_{ij}$ ($\lambda'_{ij}$) are shown.
(c) Top view. 
The shaded region indicates a beard-beard unit cell. 
(d,e) Tight binding band structure of the bilayer black phosphorus without spin-orbit coupling. 
$\Gamma$, $X$, $Y$, and $S$ denote time-reversal invariant momenta $(k_x,k_y)=(0,0)$, $(\pi/a_x,0)$, $(0,\pi/a_y)$, and $(\pi/a_x,\pi/a_y)$, respectively.   
(d) 
Blue: $E_{\rm ext}=0$ V/\AA. 
The direct band gap at $\Gamma$ is 1.12 eV. 
As $E_{\rm ext}$ is increased, the potential difference between two layers reduces
the band gap which closes at $E_{\rm ext}=$0.80 V/\AA. 
Red: $E_{\rm ext}=$1.06 V/\AA. 
The system is a 2D WSM with Weyl points along the line ${\rm Y}-\Gamma$.
The inset shows the enlarged band structure near the gap closing point.
}
\label{fig:tight-binding}
\end{figure*}

We first neglect spin-orbit coupling and treat electrons as spinless fermions.
The relevant Hamiltonian is
\begin{align}
H_{t}
=\sum_{ij,\rm intra}t_{ij}c^{\dagger}_ic_j
+\sum_{ij,\rm inter}t'_{ij}c^{\dagger}_ic_j.
\end{align}
where the intralayer hopping amplitudes $t_1= -1.220$ eV, $t_2= 3.665$ eV, $t_3= -0.205$ eV, $t_4= -0.105$ eV, $t_5= -0.055$ eV, and interlayer amplitudes $t_1'=0.295$ eV, $t_2'=0.273$ eV, $t_3'=-0.151$ eV, $t_4'=-0.091$ eV as given in Ref.~\onlinecite{Rudenko}.
The definition of hopping amplitudes is described in Fig.~\ref{fig:tight-binding}(a). 
Applied electric field induces the potential difference between layers.
For convenience, we assume uniform increase of the on-site potential along
the vertical direction. Namely, the potential difference between two sublayers
within a single monolayer and also that between neighboring monolayers are taken to be $E_{\rm eff}a_{z}/2$.
$E_{\rm eff}=E_{\rm ext}/\epsilon_r$ is the effective electric field including screening effect, where $E_{\rm ext}$ is the external electric field, and $\epsilon_r$ is the average relative permittivity.
We take the value $\epsilon_r=2.9$ following Ref.~\onlinecite{permittivity}, and assume that the same value can be used also for WSM and QSHI phases.
At the critical strength of electric field, a band inversion happens
at the $\Gamma$ point, which generates two Weyl points aligned on
the $k_{y}$ axis [See Fig.~\ref{fig:tight-binding}(d)]. The resulting Weyl points are protected
by $I_{ST}$ as discussed before. Since the Weyl points
are on the $\Gamma Y$ line, its stability can also be
understood by comparing the glide mirror 
$G_x$ eigenvalues~\cite{Zunger}. Including 
spin degrees of freedom, each gap-closing point becomes four-fold degenerate,
thus it can be considered as a Dirac point.
 
Now let us include spin-orbit coupling.
Microscopic one particle spin-orbit coupling term is $V_{so}=\frac{\hbar}{4m_e^2c^2} (\nabla V \times {\bf p}) \cdot {\bf s}$ where $\bf s$ indicates
Pauli matrices describing spin degrees of freedom.
When we construct a tight-binding Hamiltonian,
$\bm{p}$ is treated as $\bm{p}\propto -i \bm{d}_{ij}$
with ${\bf d}_{ij}={\bf r}_i-{\bf r}_j$, and
the terms associated with $\nabla_{\perp}V$ and $\nabla_{\parallel}V$
are distinguished.
The resulting Hamiltonian is
\begin{align}
H_R+H_{SO}
&= i\lambda_{R}\sum_{ij}c^{\dagger}_i ({\bf s}\times \hat{\bf d}_{ij})\cdot \hat{z}c_j\notag\\
&\quad+\sum_{k=x,y,z}i\lambda_{SO}^{k}\sum_{ij}c^{\dagger}_i\nu^{k}_{ij}s_kc_j.
\end{align}
where $\lambda^k_{SO}$ and $\nu^k_{ij}$ denote the magnitude and the sign of $\frac{-i\hbar}{4m_e^2c^2} (\nabla_{\parallel} V \times {\bf p})_k$, respectively. 
The detailed form of the Hamiltonian is constrained by $G_{x}$, $M_{y}$ and $T$ symmetries. 
Within a layer, we have one parameter $\lambda_{SO2}^y$ due to in-plane field for 2-3 and 1-4 bonds, and three parameters $\lambda_{R1}$, $\tilde{\lambda}_{R1}$, and $\lambda_{R2}$ result from the ouf-of-plane field for 1-2, 3-4, and 2-3 and 4-1 bonds, respectively.
To describe the WSM-QSHI transition, we include four interlayer spin-orbit coupling parameters $\lambda_{SO1}'^{x}$, $\lambda_{SO1}'^{y}$, $\lambda_{SO1}'^{z}$, and $\lambda_{R1}'$ resulting from the interlayer hopping between the site 1 and 4.
The magnitude of interlayer spin-orbit coupling is smaller than that of the intralayer one in the lattice model, but their effects on low-energy bands are comparable because the states associated with conduction and valence bands are localized near the bottom and top layers, respectively.
Although the inversion breaking spin-orbit coupling terms may change depending on $E_{\rm ext}$, they are treated as constants since we are focusing on the region near the critical electric field $E_{c1}$ or $E_{c2}$.

\subsection{Numerical Confirmation of the Topological Phase Transition}
\label{subsec:phosphorene_band}

\begin{figure*}[h!]
\centering
\includegraphics[width=\textwidth]{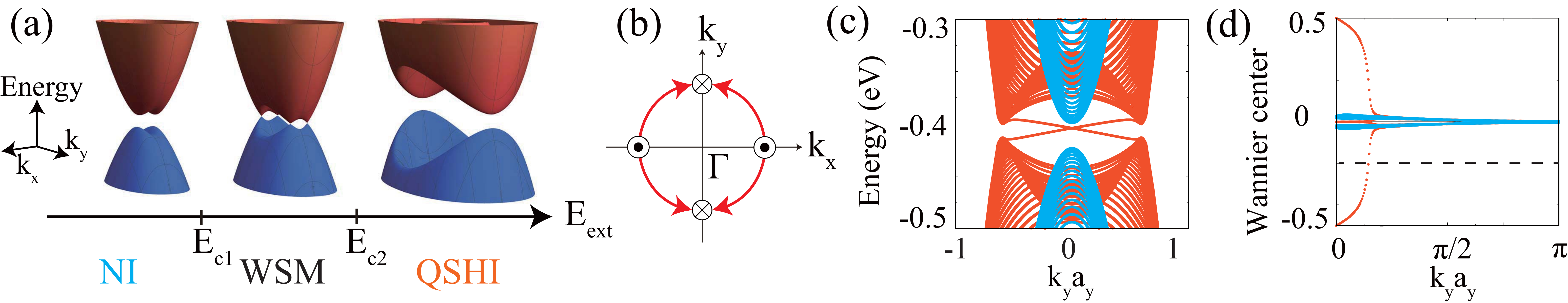}
\caption{
(a) Evolution of the band structure of a bilayer black phosphorus with spin-orbit coupling
as electric field $E_{\rm ext}$ increases. 
The band gap closes at two points on the $k_x$-axis when $E_{c1}=0.81 $ V/\AA. 
As $E_{\rm ext}$ increases further, four Weyl points are generated, and eventual pair-annihilated on the $k_y$-axis at $E_{c2}=0.85$ V/\AA.
(b) Schematic trajectories of Weyl points as $E_{\rm ext}$ increases.
Weyl points move within the range $|k|\lesssim 0.4$ in which our low energy analysis is valid.
$\odot$ and $\otimes$ are the locations where pair-creation and pair-annihilation happen.
(c,d) Confirmation of $Z_{2}$ nontriviality when $E_{\rm ext}>E_{c2}$.
The data for $E_{\rm ext}=$0.90 V/\AA (orange) are compared with those for $E_{\rm ext}=$0.80 V/\AA (blue).
(c) Energy spectrum of a finite-size system with 200 unit cells in the $x$-direction (armchair direction), while keeping translational invariance along the $y$-direction (zigzag direction).
(d) Evolution of Wannier function centers. 
The eigenvalues cross the reference line (dashed) once, indicating the $Z_2$ non-trivial nature of the gapped phase.
We used $\lambda^y_{SO2}=15$ meV, $\lambda_{R1}=12$ meV, $\tilde{\lambda}_{R1}=8$ meV, $\lambda_{R2}=5$ meV, $\lambda_{SO1}'^{x}=\lambda_{SO1}'^{y}=\lambda_{R1}'=5$ meV, and $\lambda_{SO}'^z=50$ meV here.
We set $\lambda_{SO}'^z$ to have a large value to locate Weyl points near the $\Gamma$ point.
}
\label{fig:phase_transition}
\end{figure*}

Fig.~\ref{fig:phase_transition}(a) shows the evolution of the band structure across the TPT induced by perpendicular electric field $E_{\rm ext}$.
At the critical strength of electric field $E_{c1}$, the band gap closes at two points on the $x$-axis unlike the spinless case where band gap closes only at the $\Gamma$ point. Each gap-closing point at $E_{\rm ext}=E_{c1}$ splits into two 2D Weyl points as $E_{\rm ext}$ increases further, thus the system has four Weyl points in total~\cite{Choi}.
Since $I_{ST}$ symmetry protects each Weyl point, the semimetal with four Weyl points can form a stable phase. 
Moreover, as the strength of $E_{\rm ext}$ increases further, four Weyl points approach
$k_{y}$ axis and merge pair-wise again at $E_{c2}$ leading to another gapped phase.
As shown before, the partner switching between Weyl point pairs in the intermediate semimetal phase leads to the jump of the $Z_2$ invariant.
To confirm the $Z_2$ non-triviality of the gapped phase when $E_{\rm ext}>E_{c2}$,
we compute the energy spectrum of a slab structure with finite length and also
the evolution of the Wannier function centers. 
Fig.~\ref{fig:phase_transition}(c) clearly shows the presence of helical edge states on the boundary of the slab. Moreover, Fig.~\ref{fig:phase_transition}(d)
shows that the Wannier centers cross
the reference line (the dotted line in the figure) odd number of times,
which confirms that the resulting gapped phase is a quantum spin Hall insulator.

\subsection{Critical Field vs $s_{z}$-Dependent Spin-Orbit Interaction}

\begin{figure*}[h!]
  \centering
    \includegraphics[width=\textwidth]{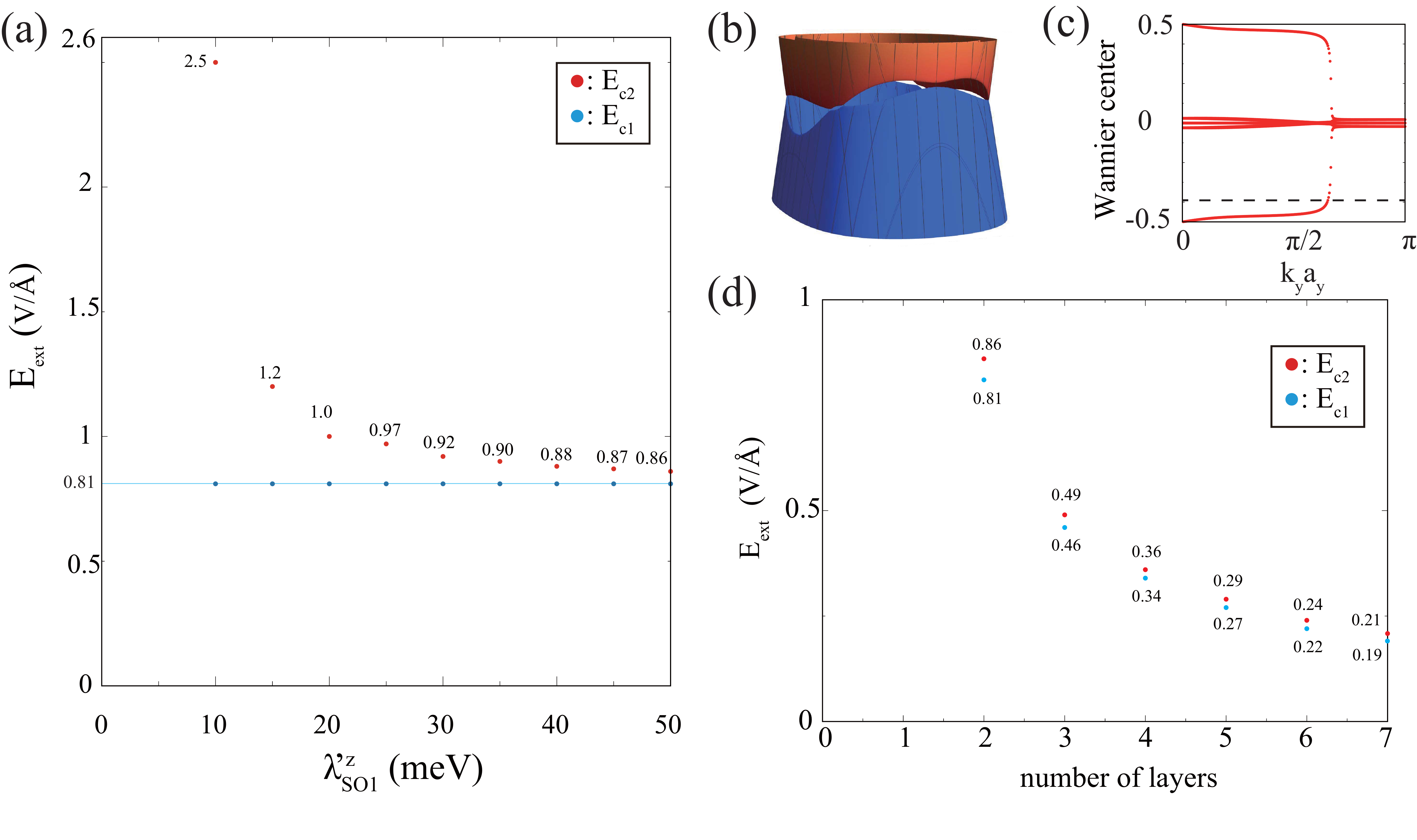}
  \caption{
(a) $\lambda_{SO}'^z$ dependence of the critical electric field $E_{c}$.
Blue (red) points indicate the critical electric field $E_{c1}$ ($E_{c2}$) at which the band gap closes (reopens) as the electric field increases.
The other spin-orbit coupling terms are assumed to have fixed values such as
$\lambda^y_{SO2}=15$ meV, $\lambda_{R1}=12$ meV, $\tilde{\lambda}_{R1}=8$ meV, $\lambda_{R2}=5$ meV, $\lambda_{SO1}'^{x}=\lambda_{SO1}'^{y}=\lambda_{R1}'=5$ meV.
As $\lambda_{SO}'^z$ becomes weaker, the critical electric field $E_{c2}$ increases while $E_{c1}$ does not change.
At around $\lambda_{SO}'^z\sim$10 meV, there is a jump in the value of $E_{c2}$ which is due to the appearance of another gap closing points at about $E_{\rm ext}=$ 1.3 V/\AA.
At $\lambda_{SO}'^z=$10 eV, the additional Weyl points are annihilated at $E_{\rm ext}=E_{c2}=2.5$ V/\AA ~ which opens the full band gap. Whereas the Weyl points created at $E=E_{c1}$ are pair-annihilated on the $k_y$-axis at $E_{\rm ext}$=1.4 V/\AA.
(b) The band structure at $\lambda_{SO}'^z=$10 meV and $E_{\rm ext}=$ 1.76 V/\AA.
(c) Evolution of Wannier function centers at $\lambda_{SO}'^z\sim$10 meV and $E_{\rm ext}=$2.66 V/\AA. 
It shows that the insulating phase is topologically nontrivial.
(d) Number of layers vs critical electric fields when
$\lambda_{SO}'^z$=50 meV.
We use the relative permittivity $\epsilon_r$=2.9, 3.5, 4.1, 4.5, 4.9, and 5.2 for $n_{\rm lay}$=2, 3, 4, 5, 6, and 7, respectively.
In both (a) and (d), when electric field becomes larger than $E_{c2}$,
additional gap-closing and associated TPT can occur since the band structure can be largely modified due to electric field.
}
\label{ec2}
\end{figure*}

Up to now, we have assumed $\lambda_{SO}'^{z}$ much larger than any other spin-orbit coupling constants since it plays a critical role to observe pair-annihilation of Weyl points.
Here we discuss what happens if $\lambda_{SO}'^{z}$ becomes smaller.
As $\lambda_{SO}'^z$ decreases while the other spin-orbit coupling terms are fixed, the critical electric field $E_{c2}$ becomes larger whereas $E_{c1}$ remains the same, which is consistent with the analysis of the low energy Hamiltonian.
[See Fig.~\ref{ec2}(a).]
Let us note that the value of $E_{c2}$ shows a discontinuous jump at around $\lambda_{SO}'^z\sim$10 meV.
The sudden jump of $E_{c2}$ is due to the appearance of additional band crossing points
in momentum space at strong electric field above $E_{\rm ext}\sim$1.3 V/\AA. [See Fig.~\ref{ec2}(b).]
In order to fully open the band gap by removing all the Weyl points in the Brillouin zone, it is necessary to apply larger electric field.
As shown in Fig.~\ref{ec2}(c), the creation and annihilation of the additional Weyl points do not change the topological property of the final insulating state.
It is also consistent with our general theory because the additional Weyl points cross a time-reversal invariant 1D line two times so that $Z_2$ invariant does not changes mod 2.

\subsection{Layer Number Dependence of the Critical Electric Field}

We examine the layer number dependence of the critical electric field by using
the same parameters for the Hamiltonian as in Sec.~\ref{sec:BP-tight-binding}.
For $n_{\rm lay}<8$, we have found a similar sequence of phase transitions between NI-WSM-QSHI. Weaker critical field is required as the layer number increases as shown in figure~\ref{ec2}(d), which happens since the band gap decreases as the number of layers increases.

On the other hand, when $n_{\rm lay}\ge 8$, although the NI-WSM transition accompanied by pair-creation of Weyl points happens as in the case with $n_{\rm lay}< 8$, the pair-annihilation of Weyl points on the $k_y$-axis does not happen. 
Instead, the Weyl points merged on the $k_y$ axis split again along the $k_{y}$ axis, thus we again observe the Weyl semimetal with four Weyl points on the $k_y$ axis.
This behavior of Weyl points can be explained by using our low energy Hamiltonian in the following way.
Suppose that the effective spin-orbit coupling parameter, $\lambda_2$, which induces the gap opening and thus are associated with $\lambda_{SO}'^{z}$, gets smaller as the number of layers increases while the other spin-orbit coupling terms are fixed.
When $\lambda_{3,4,5,6}$ get much bigger than $\lambda_{2}$, one can easily see that the gap-closing point at which Weyl points merge splits into Weyl points again.

One can understand intuitively the decrease of $\lambda_{2}$ as the layer number increases as follows. Let us note that the term $\lambda_2k_ys_z\sigma_x$ in the low energy Hamiltonian is off-diagonal with respect to the orbital degrees of freedom.
Since the states near conduction band minimum and the valence band maximum are localized at the bottom and top layers, respectively, $\lambda_2$ is large when the top and bottom layers are strongly coupled.
The effective coupling $\lambda_2$ is relatively larger for the bilayer as the coupling to the spin $z$-component comes from the interlayer hopping which directly couples the top and bottom layers, and it decreases with increasing the number of layers.
Other effective spin-orbit coupled terms other than $\lambda_1s_y\sigma_y$ and $\lambda_2k_ys_z\sigma_x$ are diagonal with respect to orbital degrees of freedom, thus their magnitude does not strongly depend on the layer number.
 

\section{Application: H\lowercase{g}T\lowercase{e}/C\lowercase{d}T\lowercase{e} Heterostructure under Uniaxial Strain}


In this section, we describe the topological phase transition of HgTe/CdTe heterostructure in the presence of uniaxial strain.
The band gap can be controlled not only by adjusting the number of HgTe layers in the HgTe/CdTe quantum well as proposed by Bernevig {\it et al}.\cite{BHZ} but also by applying strain\cite{strainHgTe}.
Since strain is a continuous parameter unlike the number of layers,  it is a good control parameter to observe the Weyl semimetal phase whose parametric region is small.

\subsection{Tight-Binding Model}

\begin{figure*}[h!]
  \centering
    \includegraphics[width=\textwidth]{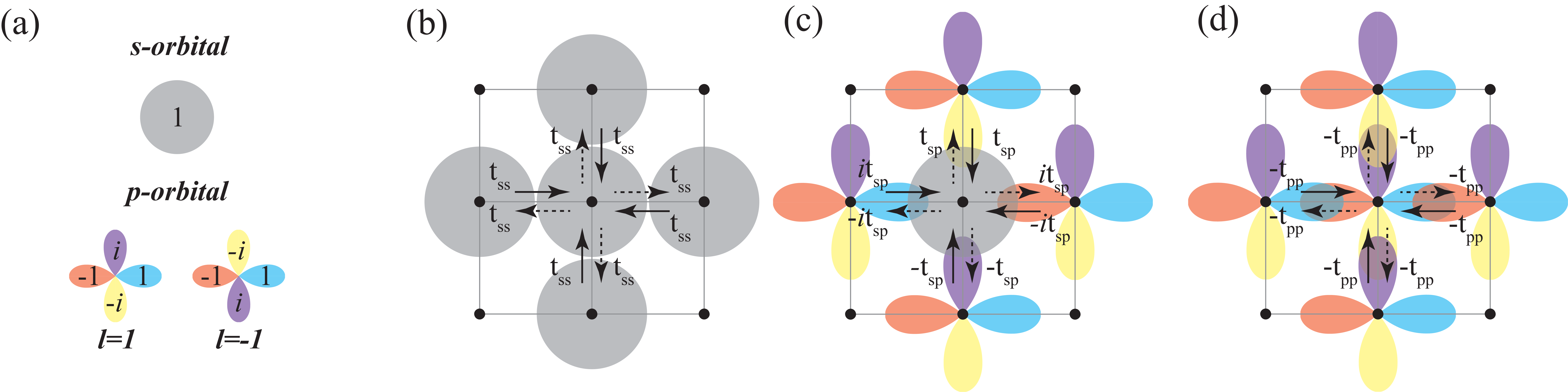}
  \caption{
Definition of hopping parameters in BHZ model.
(a) Pictorial representation of the relative phase in s-orbital and p-orbitals.
(b)-(d) Assignment of hopping amplitudes.
Relative sign of hopping amplitudes are determined by the phase difference between orbitals: $t_{ij}=\braket{i|H|j}\propto \exp\left[i(\phi_j-\phi_i)\right]$ with the phases $\phi_{i}$ and $\phi_{j}$ of states $\ket{i}$ and $\ket{j}$.
}
\label{bhz_hopping}
\end{figure*}

We consider the Bernevig-Hughes-Zhang (BHZ) tight-binding model \cite{BHZ,Fu-Kane_inversion} for HgTe/CdTe heterostructure taking into account inversion symmetry breaking.
The original BHZ model is constructed on a square lattice which is inversion symmetric.
At each site, four low energy states
\begin{align}
\ket{\psi_1}&=\ket{s,\uparrow},\notag\\
\ket{\psi_2}&=\ket{s,\downarrow},\notag\\
\ket{\psi_3}&=\ket{\frac{i}{\sqrt{2}}(p_x+ip_y),\uparrow},\notag\\
\ket{\psi_4}&=\ket{\frac{-i}{\sqrt{2}}(p_x-ip_y),\downarrow}
\end{align}
are considered, where the orbitals $s$, $p_x$, and $p_y$ are real.
The phase factors of $\psi_i$ were neglected in the main text, but we explicitly show them here because they determine hopping amplitudes.
The Hamiltonian and relevant hopping matrices will be represented by an the ordered set of four basis states.
The Hamiltonian of the system is then
\begin{align}
H=H_0+H_t=\sum_{i}c^{\dagger}_ih_0 c_i-\sum_{\braket{ij}}c^{\dagger}_it_{ij}c_j,
\end{align}
where $h_0$ is the on-site Hamiltonian
\begin{align}
h_0=
\begin{pmatrix}
\epsilon_s&0\\
0&\epsilon_p
\end{pmatrix},
\end{align}
and hopping amplitudes are defined as
\begin{align}
t_{ij}
=
\begin{pmatrix}
t_{ss}&is_zt_{sp}e^{i\theta_{ij}s_z}\\
is_zt_{sp}e^{-i\theta_{ij}s_z}&-t_{pp}\\
\end{pmatrix},
\end{align}
where $\theta_{ij}$ is the angle between the displacement vector from $j$ to $i$ and the unit vector $\hat{x}$.
The angle dependent factor is due to the spatial anisotropy of p-wave orbitals.
[See Fig.~\ref{bhz_hopping}.]

In momentum space, 
\begin{align}
h_t
&=-
\begin{pmatrix}
2t_{ss}(\cos k_x+\cos k_y)&2t_{sp}(\sin k_xs_z+i\sin k_y)\\
2t_{sp}(\sin k_xs_z-i\sin k_y)&-2t_{pp}(\cos k_x+\cos k_y)
\end{pmatrix}\notag\\
&=-(t_{ss}-t_{pp})(\cos k_x+\cos k_y)\notag\\
&\quad-(t_{ss}+t_{pp})(\cos k_x+\cos k_y)\sigma_z\notag\\
&\quad-2t_{sp}\sin k_x\sigma_xs_z+2t_{sp}\sin k_y \sigma_y,
\end{align}
where $\sigma_i$ and $s_i$ are Pauli matrices for orbital and spin degrees of freedom, respectively.
Notice that this form of Hamiltonian is the matrix transpose of that presented in Ref. \onlinecite{BHZ}.
It is because we define matrix elements of Hamiltonian as $h_{nm}=\braket{n|h|m}$ here whereas $h_{nm}=\braket{m|h|n}$ in Ref. \onlinecite{BHZ}.
The Hamiltonian $h_0$ and $h_t$ are symmetric under inversion $P=\sigma_z$.
To correctly take into account the real crystal structure of the material, we include an on-site spin-orbit coupling which breaks inversion symmetry.
\begin{align}
H_1
=\sum_{i}c^{\dagger}_ih_1 c_i
=\lambda\sum_{i}c^{\dagger}_i \sigma_ys_y c_i.
\end{align}
It is the only on-site spin-orbit coupling term which respects symmetries under
\begin{align}
T=is_yK,\quad
S_{4z}\equiv PC_3^{-1}=\frac{\textendash1-i\sigma_zs_z}{\sqrt{2}},\quad
C_{2y}=is_y.
\end{align}
In the main text, we showed the emergence of eight Weyl points in the process of phase transition from normal to topological insulating phase using the Hamiltonian $h_0+h_t+h_1$.
When $S_{4z}$ symmetry is broken, four rather than eight Weyl points will appear generically.
We now consider the effect of strain to break $S_{4z}$ symmetry down to $C_{2z}=S_{4z}^2$ symmetry.
The Hamiltonian in momentum space is given by
\begin{align}
h
&=h_0+h_t+h_1\notag\\
&=
\mu
+\left(M+2B_1(\cos k_x-1)+2B_2(\cos k_y-1)\right)\sigma_z\notag\\
&\quad
+A_1 \sin k_x\sigma_xs_z- A_2 \sin k_y \sigma_y
+\lambda \sigma_ys_y,
\end{align}
where
\begin{align}
\mu
&=\frac{1}{2}(\epsilon_{s}+\epsilon_{p})-(t^x_{ss}-t^x_{pp})\cos k_x -(t^y_{ss}-t^y_{pp})\cos k_y,\notag\\
M
&=\frac{1}{2}(\epsilon_{s}-\epsilon_{p})-(t^x_{ss}+t^x_{pp})-(t^y_{ss}+t^y_{pp}),\notag\\
B_1
&=-\frac{1}{2}(t^x_{ss}+t^x_{pp}),\notag\\
B_2
&=-\frac{1}{2}(t^y_{ss}+t^y_{pp}),\notag\\
A_1
&=-2t_{sp}^x,\notag\\
A_2
&=-2t_{sp}^y.
\end{align}

\subsection{Low-Energy Hamiltonian}

Near the $\Gamma$ point, the effective Hamiltonian up to quadratic order in momentum is
\begin{align}
h_{\rm eff}
&=
\mu
+\left(M-B_1k_x^2-B_2k_y^2\right)\sigma_z\notag\\
&\quad
+A_1 k_x\sigma_xs_z- A_2 k_y \sigma_y
+\lambda\sigma_ys_y
\end{align}
The energy spectrum is given by
\begin{align}
E^{(\pm)}_{\pm}(k_x,k_y)
&=\mu(k_x,k_y)\pm\sqrt{\left(f(k_x,k_y)\right)^2+\left(g_{\pm}(k_x,k_y)\right)^2},\notag\\
f(k_x,k_y)
&=M-B_1k_x^2-B_2k_y^2,\notag\\
g_{\pm}(k_x,k_y)
&=\sqrt{A_1^2k_x^2+A_2^2k_y^2}(\pm)|\lambda|
\end{align}
Let us assume that $M$ can be taken as a control parameter.
Band gap can be closed at four points if $A_1B_2-A_2B_1\ne 0$ as $M$ is varied.
The critical values of $M$ are
\begin{align}
M_{c_1}=B_1\left(\frac{\lambda}{A_{1}}\right)^2,\quad
M_{c_2}=B_2\left(\frac{\lambda}{A_{2}}\right)^2,
\end{align}
at which we have two anisotropic Weyl points at $(k_x,k_y)=(\pm \lambda/A_1,0)$ and $(k_x,k_y)=(\pm \lambda/A_2,0)$, respectively.

\subsection{Numerical Calculations}

\begin{figure*}[h!]
  \centering
    \includegraphics[width=\textwidth]{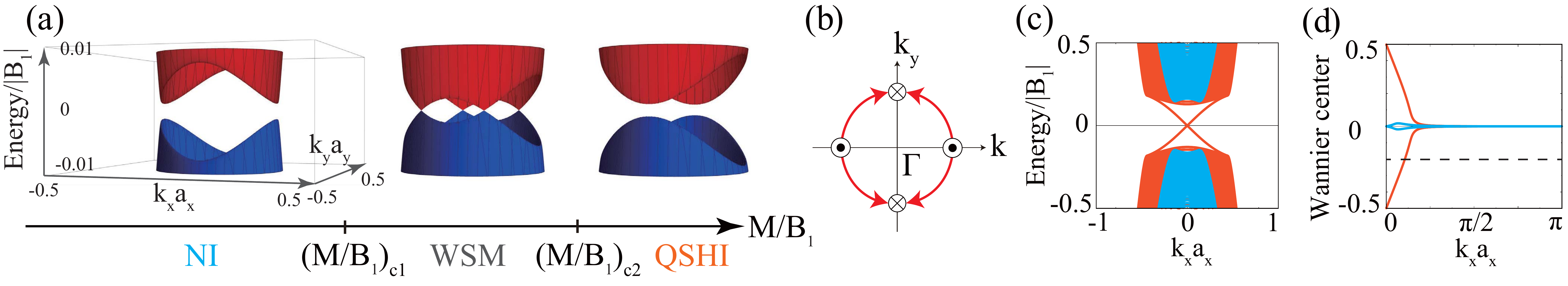}
  \caption{
  (a) Phase diagram for the TPT in modified BHZ model including uniaxial strain.
$\mu=0$, $B_2=1.1B_1$, $A_1=0.2B_1$, $A_2=0.18B_1$, $\lambda=0.05B_1$ are used here, and also in (c-d). 
The representative band structures are calculated at $M/B_1=0.1$, $0.08$ and $0.06$, respectively.
(b) Schematic diagram for the trajectory of Weyl points.
(c-d) Finite-size band structure and the evolution of Wannier function centers. 
Blue and orange lines are plotted at $M/B_1=0$ and $M/B_1=0.2$, respectively. 
Helical edge states and the partner switching betwenn Wannier functions are the evidence for the QSHI phase.
}
\label{bhz_phase}
\end{figure*}

Taking $M$ as a control parameter, we have found the phase diagram composed of the QSHI-WSM-NI [Fig.~\ref{bhz_phase}] as in the main text. 
In the semimetallic phase, four rather than eight Weyl points appear due to the absence of $S_{4z}$ symmetry.
\\\\\\\\\\\\\\\\\\\\\\\\\\\\\\\\\\\\\\\\\\\\\\\\\\\\\\\\\\\\\\\\\\\\\\

\begin{thebibliography}{99}

\bibitem{Hasan-Kane}
M. Z. Hasan and C. L. Kane
\rmp {\bf 82}, 3045 (2010).

\bibitem{Qi-Zhang}
X. -L. Qi and S. -C. Zhang
\rmp {\bf 83}, 1057 (2011).

\bibitem{Ando-Fu}
Y. Ando and L. Fu
Annu. Rev. Condens. Matter Phys. {\bf 6}, 361 (2015).

\bibitem{TCI}
T. H. Hsieh, H. Lin, J. Liu, W. Duan, A. Bansil, and L. Fu, Nat.
Commun. 3, 982 (2012);
J. C. Y. Teo, L. Fu, and C. L. Kane, Phys. Rev. B 78, 045426
(2008);
Y. Tanaka et al., Nat. Phys. 8, 800 (2012);
P. Dziawa et al., Nat. Mater. 11, 1023 (2012);
S. Xu et al., Nat. Commun. 3, 1192 (2012).

\bibitem{Murakami}
S. Murakami and S. -i. Kuga,
\prb {\bf 78}, 165313 (2008);
S. Murakami, New J. Phys. {\bf 9}, 356 (2007).

\bibitem{Murakami_2D}
S. Murakami, S. Iso, Y. Avishai, M. Onoda, and N. Nagaosa,
\prb {\bf 76}, 205304 (2007).


\bibitem{notation}
In this letter we call the linearly dispersing two-fold (four-fold) degenerate gapless point nodes as Weyl (Dirac) points. In high-energy physics, the terms Weyl and Dirac are used to name spinor representations of the Lorentz group. Dirac representation is the irreducible representation of Clifford algebra $\{\Gamma_0,\Gamma_1,...,\Gamma_d\}$ with $\{\Gamma^{\mu},\Gamma^{\nu}\}=2{\rm diag}(-1,1,...,1)$, and Weyl representation is given by a restriction of it to states with fixed eigenvalue +1 or -1 under the action of the chirality operator $\Gamma_{d+1}\equiv i^{-(d-1)/2}\Gamma_0\Gamma_1...\Gamma_d$ which can be defined only for even spacetime dimensions. Dirac representation in (2+1)D is a two component representation, and there is no Weyl representation in this sense. The denomination we use are analogous to the massless spinor representation of the 3+1d Lorentz group.


\bibitem{spacetime}
C. Fang and L. Fu,
\prb {\bf 91}, 161105(R) (2015).


\bibitem{BIA_Dai}
X. Dai, T. L. Hughes, X.-L. Qi, Z. Fang, and S. C. Zhang,
\prb {\bf 77}, 125319 (2008).

\bibitem{BIA_Konig}
M. König, H. Buhmann, L. W. Molenkamp, T. L. Hughes, C.-X. Liu, X. L. Qi, and S. C. Zhang, J. Phys. Soc. Jpn.  {\bf 77}, 031007 (2008).


\bibitem{BIA_Winkler}
R. Winkler, L. Y. Wang, Y. H. Lin, and C. S. Chu, 
Solid State Commun. {\bf 152}, 2096 (2012).



\bibitem{BIA_Zunger}
S. A. Tarasenko, M. V. Durnev, M. O. Nestoklon, E. L. Ivchenko, Jun-Wei Luo, and Alex Zunger
\prb {\bf 91}, 081302(R) (2015).


\bibitem{Kane-Mele}
C. L. Kane and E. J. Mele,
\prl {\bf 95}, 226801 (2005).


\bibitem{Fu-Kane}
L. Fu and C. L. Kane,
\prb {\bf 74}, 195312 (2006).

\bibitem{Wannier1}
B. I. Blount, 
Solid State Phys. {\bf 13}, 305 (1962).

\bibitem{Wannier2}
J. Zak,
\prl {\bf 62}, 2747 (1989).

\bibitem{supp}
See Supplemental Material at http://link.aps.org/
supplemental/10.1103/PhysRevLett.xxx.xxxxxx, 
which includes Refs. \cite{Moore-Balents,Hasegawa,YWSon,Rudenko,permittivity,strainHgTe}


\bibitem{Moore-Balents}
J. E.Moore and L. Balents,
\prb {\bf 75} 121306(R) (2007).

\bibitem{Hasegawa}
Y. Hasegawa, R. Konno, H. Nakano, and M. Kohmoto
\prb {\bf 74}, 033413 (2006).

\bibitem{YWSon}
S. -M. Choi, S. -H. Jhi, and Y. -W. Son
\prb {\bf 81}, 081407(R) (2010).

\bibitem{Rudenko} 
A. N. Rudenko and M. I. Katsnelson, 
\prb {\bf 89}, 201408(R) (2014).

\bibitem{permittivity}
P. Kumar, B. S. Bhadoria, S. Kumar, S. Bhowmick, Y. S. Chauhan, and A. Agarwal, 
\prb {\bf 93}, 195428 (2016).

\bibitem{strainHgTe}
P. Leubner, L. Lunczer, C. Br\"une, H. Buhmann, and L. W. Molenkamp
\prl {\bf 117}, 086403 (2016).


\bibitem{mirror_insulator}
A. Lau, J. van den Brink, and C. Ortix,
\prb {\bf 94}, 165164 (2016).

\bibitem{BHZ}
B. A. Bernevig, T. L. Hughes, S.-C. Zhang,
Science {\bf 314}, 1757 (2006).

\bibitem{BHZexp}
M. K\"onig, S. Wiedmann, C. Br\"une, A. Roth, H. Buhmann, L. W. Molenkamp, X.-L. Qi, S.-C. Zhang,
Science {\bf 318}, 766 (2006).




\bibitem{Fu-Kane_inversion}
L. Fu and C. L. Kane,
\prb {\bf 76}, 045302 (2007).

\bibitem{wilson_loop}
R. Yu, X. L. Qi, A. Bernevig, Z. Fang, and X. Dai,
\prb {\bf 84}, 075119 (2011).



\bibitem{phosphorene}
H. Liu, A. T. Neal, Z. Zhu, Z. Luo, X. Xu, D. Tom\'anek, and P. D. Ye,
ACS Nano {\bf 8} 4033 (2014).




\bibitem{Zunger} 
Q. Liu, X. Zhang, L. B. Abdalla, A. Fazzio, and A. Zunger, 
Nano. Lett. {\bf 15}, 1222-1228 (2015).


\bibitem{Zhang}
T. Zhang, J.-H. Lin, Y.-M. Yu, X.-R. Chen, and W.-M. Liu,
Sci. Rep. {\bf 5}, 13927 (2015).

\bibitem{Sisakht}
E. Taghizadeh Sisakht, F. Fazileh, M. H. Zare, M. Zarenia, and F. M. Peeters,
\prb {\bf 94}, 085417 (2016).



\bibitem{KSKim}
J. Kim, S. S. Baik, S. H. Ryu, Y. Sohn, S. Park, B.-G. Park, J. Denlinger, Y. Yi, H. J. Choi, K. S. Kim,
Science, {\bf 349}, 723 (2015).

\bibitem{Choi} 
S. S. Baik, K. S. Kim, Y. Yi, and H. J. Choi, 
Nano. Lett. {\bf 15} 7788-7793 (2015).

\bibitem{2Dmaterial}
G. Zheng, Y. Jia, S. Gao, and S.-H. Ke, \prb {\bf 94}, 155448 (2016);
C. Kamal and M. Ezawa, \prb {\bf 91}, 085423 (2015);
S. Zhang, M. Xie, F. Li, Z. Yan, Y. Li, E. Kan, W. Liu, Z. Chen, and H. Zeng,
Angew. Chem. Int. Ed. {\bf 55}, 1666 (2016);
T. Nagao, J. T. Sadowski, M. Saito, S. Yaginuma, Y. Fujikawa, T. Kogure, T. Ohno, Y. Hasegawa, S. Hasegawa, and T. Sakurai,
\prl {\bf 93}, 105501 (2004).

\bibitem{dumbbell}
X. Chen, L. Li, and M. Zhao,
RSC adv. {\bf 5}, 72462 (2015);
P. Tang, P. Chen, W. Cao, H. Huang, S. Cahangirov, L. Xian, Y. Xu, S.-C. Zhang, W. Duan, and A. Rubio,
\prb {\bf 90}, 121408(R) (2014).

\bibitem{monobromide}
J.-J. Zhou, W. Feng, C.-C. Liu, S. Guan, and Y. Yao,
Nano Lett. {\bf 14}, 4767 (2014).

\bibitem{interaction}
D. I. Pikulin and T. Hyart, 
\prl {\bf 112}, 176403 (2014).


\bibitem{RG-largeN}
H. Isobe, B. -J. Yang, A. Chubukov, J. Schmalian, N. Nagaosa,
\prl {\bf 116}, 076803 (2016).

\bibitem{disorder}
D. Carpentier, A. A. Fedorenko, and E. Orignac, 
EPL {\bf 102}, 67010 (2013).


\end{thebibliography}
\end{document}